\documentclass{aa}
\usepackage[dvips]{graphicx}
\usepackage{indentfirst}
\usepackage{amssymb}
\usepackage{txfonts}
\usepackage{natbib}
\begin{document}
\title{Soft band X/K luminosity ratios for gas-poor early-type galaxies}
\author{\'A. Bogd\'an \inst{1}
\and M. Gilfanov\inst{1,2}}

\offprints{bogdan@mpa-garching.mpg.de (\'AB); gilfanov@mpa-garching.mpg.de 
  (MG)}

\institute{\inst{1} Max-Planck-Institut f\"ur Astrophysik, Karl- 
Schwarzschild-Str. 1, 85741 Garching bei M\"unchen, Germany \\
            \inst{2} Space Research Institute, Russian Academy of  
Sciences, Profsoyuznaya 84/32, 117997 Moscow, Russia}

\date{}

\abstract
{}
{
We aim to place upper limits on the combined X-ray emission from the  
population of steady nuclear-burning white dwarfs  in galaxies. In the  
framework of the single-degenerate scenario, these systems, known as supersoft 
sources, are believed to be likely progenitors of Type Ia  
supernovae.
}
{
From the \textit{Chandra} archive, we selected normal early-type galaxies with   
the point source detection sensitivity better than $10^{37} \ \mathrm{erg
\ s^{-1}}  $  in order to minimize the contribution of unresolved 
low-mass X-ray binaries. The galaxies, contaminated by emission from 
ionized ISM, were identified based on the analysis of radial surface 
brightness profiles and energy spectra. The sample was complemented 
by the bulge of M31 and the data for the solar neighborhood.    
To cover a broad range of ages, we also included   NGC3377 and 
NGC3585 which represent the  young end of the age distribution for 
elliptical galaxies. Our final sample includes eight gas-poor galaxies 
for which we determine $L_X/L_K$ ratios in the $0.3-0.7$ keV energy 
band. This choice of the energy band was optimized to detect soft 
emission from thermonuclear-burning on the  surface of an accreting 
white dwarf. In computing the $L_X$ we included both unresolved emission 
and soft resolved sources with the color temperature of 
$kT_{bb}\le 200$ eV.}
{We find that the X/K luminosity ratios are in a rather narrow range  
of $(1.7-3.2)\cdot 10^{27} \ \mathrm{erg \ s^{-1} \ L_{K,\odot}}$. The 
data show no obvious trends with mass, age, or metallicity of the host  
galaxy, although a weak anti-correlation with the Galactic NH appears  
to exist. It is much flatter than predicted for a blackbody emission  
spectrum with temperature of $\sim 50-75$ eV, suggesting that  
sources with such soft spectra contribute significantly less than a
half to the observed X/K ratios. However, the correlation of the  
X/K ratios with NH  has a significant scatter and in the strict  
statistical sense cannot be adequately described by a superposition  
of a power law and a blackbody components with reasonable parameters,  
thus precluding quantitative constraints on the contribution from soft  
sources.
}
{}


\maketitle
\section{Introduction}

In the framework of the singe-degenerate scenario \citep{whelan}, 
white dwarfs (WDs) accreting from a donor star in a binary 
system and steadily burning  the accreted material on 
their surface are believed to be a  likely path to the 
Type Ia supernova \citep{hillebrandt,livio}. Nuclear burning 
is only stable (required for the WD to grow in mass) if the  mass 
accretion rate is high enough,  $\dot{M}\sim 10^{-7}-10^{-6} M_\odot$/yr.
Given that the nuclear-burning efficiency for hydrogen is  $\epsilon_H  
\approx 6 \cdot 10^{18}$ erg/g, the bolometric luminosity of such  
systems are in the $ 10^{37}-10^{38} \ \mathrm{erg
\ s^{-1}} $  range, potentially making them  bright X-ray sources. 
Their emission, however, has a rather low effective temperature, 
$T_{\mathrm{eff}}\lesssim 50-100$ eV so is 
prone to absorption by cold ISM \citep{nature}. The  
brightest and hardest sources of this type are indeed observed as  
supersoft sources in the Milky Way and nearby galaxies \citep{greiner}. 
The rest of the population, however, remains unresolved -- weakened 
by absorption and blended with other types of faint X-ray sources 
-- thus makes its contribution to the unresolved X-ray emission 
from galaxies.

\begin{table*}[!ht]
\caption{The list of early-type galaxies and galaxy bulges studied in this paper.}
\begin{minipage}{18cm}
\renewcommand{\arraystretch}{1.3}
\centering
\begin{tabular}{c c c c c c c c c c}
\hline 
Name & Distance &   $N_{H}$ & Morphological & $ L_{K}$  & $ T_{\mathrm{obs}} $ & $ T_{\mathrm{filt}} $ & $L_{\mathrm{lim}}$  &  $ R_{1} $ & $ R_{2} $ \\ 
     & (Mpc)    & (cm$^{-2}$) & type        &($\mathrm{L_{K,\odot}}) $ & (ks)  & (ks)                  &  ($ \mathrm{erg \ s^{-1}} $) & ($\arcsec$) & ($\arcsec$) \\ 
     &   (1)    &    (2)      &     (3)     &      (4)                 &   (5) &  (6)                  &   (7)                        &     (8)     & (9) \\
\hline 
M31 bulge   & $ 0.78^a   $ & $ 6.7 \cdot 10^{20} $ & SA(s)b         & $ 3.7 \cdot 10^{10} $ & $ 180.1 $ & $ 143.7 $   & $ 2 \cdot 10^{35} $ & $ 300 $ & $ 720 $  \\   
M32         & $ 0.805^b  $ & $ 6.3 \cdot 10^{20} $ & cE2            & $ 8.5 \cdot 10^{8} $ & $ 178.8 $ & $ 172.2 $    & $ 1 \cdot 10^{34} $ & $  25 $ & $  90 $  \\   
M60         & $ 16.8^c   $ & $ 2.2 \cdot 10^{20} $ & E2             & $ 2.2 \cdot 10^{11} $ & $ 109.3 $ & $  90.5 $   & $ 9 \cdot 10^{36} $ & $  50 $ & $ 125 $  \\
M84         & $ 18.4^c   $ & $ 2.6 \cdot 10^{20} $ & E1             & $ 1.6 \cdot 10^{11} $ & $ 117.0 $ & $ 113.8 $   & $ 9 \cdot 10^{36} $ & $  50 $ & $ 100 $  \\
M105        & $ 9.8^d    $ & $ 2.8 \cdot 10^{20} $ & E1             & $ 4.1 \cdot 10^{10} $ & $ 341.4 $ & $ 314.0 $   & $ 1 \cdot 10^{36} $ & $  30 $ & $  90 $  \\
NGC1291     & $ 8.9^e    $ & $ 2.1 \cdot 10^{20} $ & (R)SB(s)0/a    & $ 6.3 \cdot 10^{10} $ & $ 76.7 $  & $ 51.2 $    & $ 5 \cdot 10^{36} $ & $  60 $ & $ 130 $  \\
NGC3377     & $ 11.2^c   $ & $ 2.9 \cdot 10^{20} $ & E5-6           & $ 2.0 \cdot 10^{10} $ & $ 40.2 $  & $ 34.0 $    & $ 2 \cdot 10^{37} $ & $   - $ & $  79 $  \\
NGC3585     & $ 20.0^c   $ & $ 5.6 \cdot 10^{20} $ & E7/S0          & $ 1.5 \cdot 10^{11} $ & $ 95.9 $  & $ 89.1 $    & $ 2 \cdot 10^{37} $ & $  50 $ & $ 180 $  \\
NGC4278     & $ 16.1^c   $ & $ 1.8 \cdot 10^{20} $ & E1-2           & $ 5.5 \cdot 10^{10} $ & $ 467.7 $ & $ 443.0 $   & $ 2 \cdot 10^{36} $ & $  50 $ & $ 110 $  \\ 
NGC4365     & $ 20.4^c   $ & $ 1.6 \cdot 10^{20} $ & E3             & $ 1.1 \cdot 10^{11} $ & $ 198.3 $ & $ 181.4 $   & $ 8 \cdot 10^{36} $ & $  30 $ & $  80 $  \\
NGC4636     & $ 14.7^c   $ & $ 1.8 \cdot 10^{20} $ & E/S0\_1        & $ 8.1 \cdot 10^{10} $ & $ 212.5 $ & $ 202.1 $   & $ 5 \cdot 10^{36} $ & $  30 $ & $ 102 $  \\
NGC4697     & $ 11.8^c   $ & $ 2.1 \cdot 10^{20} $ & E6             & $ 5.1 \cdot 10^{10} $ & $ 195.6 $ & $ 162.0 $   & $ 4 \cdot 10^{36} $ & $  30 $ & $  95 $  \\
NGC5128     & $ 3.7^f    $ & $ 8.6 \cdot 10^{20} $ & S0 pec         & $ 5.8 \cdot 10^{10} $ & $ 569.1 $ & $ 568.9 $   & $ 1 \cdot 10^{36} $ & $  65 $ & $ 240 $  \\
Sagittarius & $ 0.0248^g $ & $ 1.2 \cdot 10^{21} $ & dSph(t)        & $ 8.9 \cdot 10^{5}  $ & $  30.1 $ & $  29.7 $   & $ 1 \cdot 10^{32} $ & $   - $ & $  55 $  \\ 
\hline \\
\end{tabular} 
\end{minipage}
(1) References for distances:  $^a$ \citet{stanek,macri} -- $^b$ \citet{m32distance} -- $^c$ \citet{ngc4278distance} -- $^d $ \citet{m105distance} -- $^e$ \citet{ngc1291distance} -- $^f$ \citet{ngc5128distance} -- $^g$ \citet{sagittariusdistance} (2) Galactic absorption column density \citep{dickey} (3) Taken from NED (http://nedwww.ipac.caltech.edu/) (4) K-band luminosity of galaxies (5) and (6) Exposure times before and after data filtering (7) Point source detection sensitivity in the $ 0.5-8 $ keV energy range (8) and (9) The radii used for the inner and outer regions in obtaining the spectra presented in Fig. \ref{fig:spectra}. The orientation and shape of the regions were taken from K-band measurements (http://irsa.ipac.caltech.edu/applications/2MASS/). \\
\label{tab:list2}
\end{table*}  

We have recently proposed that the combined energy output of  
accreting WDs can be used to measure the rate at which WDs 
increase their mass in galaxies \citep{nature}. This allowed us  
to severely constrain the contribution of the single-degenerate scenario  
to the observed Type Ia supernova rate in early-type galaxies. The critical  
quantity in our argument is the X-ray to K-band luminosity ratio of  
the population of accreting white dwarfs. This quantity cannot be  
measured unambiguously for several reasons.  First, galaxies  
have large populations of bright compact X-ray sources, -- accreting  
neutron stars and black holes in binary systems (Gilfanov, 2004).  
Although their spectra  are relatively hard, these sources make a  
significant contribution to  X/K ratios, even in the soft band. Unless  
their contribution is removed,  the obtained X/K ratios are rendered  
useless. This requires  adequate sensitivity and angular resolution, 
a combination of qualities that currently can only be delivered  
by \textit{Chandra} observatory. Another source of contamination is the
hot ionized gas present in some of galaxies \citep{mathews}. Although 
there is a general correlation between the gas luminosity and the mass of 
the galaxy, the large dispersion  precludes an accurate subtraction 
of the gas contribution based on, for example, optical properties of
galaxies. The gas contribution may increase the X/K ratio by $\sim 1-2$ 
orders of magnitude, therefore  gas-rich galaxies need to be identified 
and excluded from the sample. Finally, other types of faint sources do 
exist and contribute to the unresolved X-ray emission. Only upper limits 
on the luminosity of WDs can be obtained, because different components 
in the  unresolved emission cannot be separated. 

The aim of this paper is to measure $L_X/L_K$ 
ratios in the $0.3 - 0.7 $ keV  band for a sample of nearby gas-poor 
galaxies. The energy range has been optimized  to detect emission
from nuclear-burning white dwarfs, considering the range of 
effective temperatures, absorption column densities, and the effective 
area curve of \textit{Chandra} detectors.

The paper is structured as follows: in Sect. 2 we describe the  
sample selection, the data preparation, and its analysis. We identify and  
remove gas-rich galaxies from the sample in Sect. 3. The  
obtained X/K ratios are presented and discussed in Sect. 4.  Our  
results are summarized in Sect. 5.

\begin{table*}[!ht]
\caption{The list of \textit{Chandra} observations used in this paper.}
\begin{minipage}{18cm}
\centering
\begin{tabular}{c c c c c c || c c c c c c}
\hline \\
Obs-ID & $ T_{\mathrm{obs}} $, ks &  $ T_{\mathrm{filt}} $, ks & Date & Det.$^\star$ & Galaxy & Obs-ID & $ T_{\mathrm{obs}} $, ks &  $ T_{\mathrm{filt}} $, ks & Date & Det.$^\star$ &Galaxy \\ \\
\hline \\
$ 303  $ & $ 12.0 $ & $ 8.2   $ &  1999 Oct 13 & I  & M31    &      $ 7075 $ & $ 84.2 $ & $ 79.0  $ &  2006 Jul 03 & S  &  M105       \\
$ 305  $ & $ 4.2  $ & $ 4.0   $ &  1999 Dec 11 & I  & M31    &      $ 7076 $ & $ 70.1 $ & $  66.7 $ &  2007 Jan 10 & S  &  M105       \\
$ 306  $ & $ 4.2  $ & $ 4.1   $ &  1999 Dec 27 & I  & M31    &      $  795 $ & $  39.7$ & $ 31.7  $ &  2000 Jun 27 & S  &  NGC1291    \\
$ 307  $ & $ 4.2  $ & $ 3.1   $ &  2000 Jan 29 & I  & M31    &      $ 2059 $ & $  37.0$ & $ 19.5  $ &  2000 Nov 07 & S  &  NGC1291    \\
$ 308  $ & $ 4.1  $ & $ 3.7   $ &  2000 Feb 16 & I  & M31    &      $ 2934 $ & $  40.2$ & $ 34.0  $ &  2003 Jan 06 & S  &  NGC3377    \\
$ 311  $ & $ 5.0  $ & $ 3.9   $ &  2000 Jul 29 & I  & M31    &      $ 2078 $ & $  35.8$ & $ 33.1  $ &  2001 Jun 03 & S  &  NGC3585    \\
$ 312  $ & $ 4.7  $ & $ 3.8   $ &  2000 Aug 27 & I  & M31    &      $ 9506 $ & $  60.2$ & $ 56.0  $ &  2008 Mar 11 & S  &  NGC3585    \\
$ 1575 $ & $ 38.2 $ & $38.2   $ &  2001 Oct 05 & S  & M31    &      $ 4741 $ & $ 37.9 $ & $ 34.7  $ &  2005 Feb 03 & S  &  NGC4278    \\
$ 1577 $ & $ 5.0  $ & $ 4.9   $ &  2001 Aug 31 & I  & M31    &      $ 7077 $ & $ 111.7$ & $ 103.2 $ &  2006 Mar 16 & S  &  NGC4278    \\
$ 1583 $ & $ 5.0  $ & $ 4.1   $ &  2001 Jun 10 & I  & M31    &      $ 7078 $ & $ 52.1 $ & $ 46.4  $ &  2006 Jul 25 & S  &  NGC4278    \\
$ 1585 $ & $ 5.0  $ & $ 4.1   $ &  2001 Nov 19 & I  & M31    &      $ 7079 $ & $ 106.4$ & $ 100.2 $ &  2006 Oct 24 & S  &  NGC4278    \\
$ 2895 $ & $ 4.9  $ & $ 3.2   $ &  2001 Dec 07 & I  & M31    &      $ 7080 $ & $ 56.5 $ & $  53.6 $ &  2007 Apr 20 & S  &  NGC4278    \\
$ 2896 $ & $ 5.0  $ & $ 3.7   $ &  2002 Feb 06 & I  & M31    &      $ 7081 $ & $ 112.1$ & $ 104.9 $ &  2007 Feb 20 & S  &  NGC4278    \\
$ 2897 $ & $ 5.0  $ & $ 4.1   $ &  2002 Jan 08 & I  & M31    &      $ 2015 $ & $  41.0$ & $ 36.1  $ &  2001 Jun 02 & S  &  NGC4365    \\
$ 2898 $ & $ 5.0  $ & $ 3.2   $ &  2002 Jun 02 & I  & M31    &      $ 5921 $ & $  40.0$ & $ 37.3  $ &  2005 Apr 28 & S  &  NGC4365    \\
$ 4360 $ & $ 5.0  $ & $ 3.4   $ &  2002 Aug 11 & I  & M31    &      $ 5922 $ & $  40.0$ & $ 38.3  $ &  2005 May 09 & S  &  NGC4365    \\
$ 4678 $ & $ 4.9  $ & $ 2.7   $ &  2003 Nov 09 & I  & M31    &      $ 5923 $ & $  40.1$ & $ 34.7  $ &  2005 Jun 14 & S  &  NGC4365    \\
$ 4679 $ & $ 4.8  $ & $ 2.7   $ &  2003 Nov 26 & I  & M31    &      $ 5924 $ & $  27.1$ & $ 25.7  $ &  2005 Nov 25 & S  &  NGC4365    \\
$ 4680 $ & $ 5.2  $ & $ 3.2   $ &  2003 Dec 27 & I  & M31    &      $ 7224 $ & $  10.1$ & $ 9.3   $ &  2005 Nov 26 & S  &  NGC4365    \\
$ 4681 $ & $ 5.1  $ & $ 3.3   $ &  2004 Jan 31 & I  & M31    &      $  323 $ & $  53.0$ & $ 46.4  $ &  2000 Jan 26 & S  &  NGC4636    \\
$ 4682 $ & $ 4.9  $ & $ 1.2   $ &  2004 May 23 & I  & M31    &      $  324 $ & $   8.5$ & $  4.7  $ &  1999 Dec 04 & I  &  NGC4636    \\
$ 7064 $ & $ 29.1 $ & $ 23.2  $ &  2006 Dec 04 & I  & M31    &      $ 3926 $ & $  75.7$ & $ 75.7  $ &  2003 Feb 14 & I  &  NGC4636    \\
$ 7068 $ & $  9.6 $ & $  7.7  $ &  2007 Jun 02 & I  & M31    &      $ 4415 $ & $  75.3$ & $ 75.3  $ &  2003 Feb 15 & I  &  NGC4636    \\
$ 2017 $ & $ 46.5 $ & $ 42.2  $ &  2001 Jul 24 & S  & M32    &      $ 784  $ & $  39.8$ & $ 37.8  $ &  2000 Jan 15 & S  &  NGC4697    \\
$ 2494 $ & $ 16.2 $ & $ 13.9  $ &  2001 Jul 28 & S  & M32    &      $ 4727 $ & $  40.4$ & $ 36.4  $ &  2003 Dec 26 & S  &  NGC4697    \\
$ 5690 $ & $ 116.1$ & $ 116.1 $ &  2005 May 27 & S  & M32    &      $ 4728 $ & $  36.2$ & $ 31.4  $ &  2004 Jan 06 & S  &  NGC4697    \\
$ 785  $ & $ 38.6 $ & $ 23.2  $ &  2000 Apr 20 & S  & M60    &      $ 4729 $ & $  38.6$ & $ 19.5  $ &  2004 Feb 12 & S  &  NGC4697    \\
$ 8182 $ & $ 53.0 $ & $ 49.6  $ &  2007 Jan 30 & S  & M60    &      $ 4730 $ & $  40.6$ & $ 36.9  $ &  2004 Aug 18 & S  &  NGC4697    \\
$ 8507 $ & $ 17.7 $ & $ 17.7  $ &  2007 Feb 01 & S  & M60    &      $ 7797 $ & $  98.2$ & $ 98.0  $ &  2007 Mar 22 & I  &  NGC5128    \\
$ 803  $ & $ 28.8 $ & $ 28.6  $ &  2000 May 19 & S  & M84    &      $ 7798 $ & $  92.0$ & $ 92.0  $ &  2007 Mar 27 & I  &  NGC5128    \\
$ 5908 $ & $ 46.7 $ & $ 46.1  $ &  2005 May 01 & S  & M84    &      $ 7799 $ & $  96.0$ & $ 96.0  $ &  2009 Mar 30 & I  &  NGC5128    \\
$ 6131 $ & $ 41.5 $ & $ 39.1  $ &  2005 Nov 07 & S  & M84    &      $ 7800 $ & $  92.0$ & $ 92.0  $ &  2007 Apr 17 & I  &  NGC5128    \\
$ 1587 $ & $ 31.9 $ & $ 25.1  $ &  2001 Feb 13 & S  & M105   &      $ 8489 $ & $  95.2$ & $ 95.2  $ &  2007 May 08 & I  &  NGC5128    \\
$ 7073 $ & $ 85.2 $ & $ 76.8  $ &  2006 Jan 23 & S  & M105   &      $ 8490 $ & $  95.7$ & $ 95.7  $ &  2007 May 30 & I  &  NGC5128    \\
$ 7074 $ & $ 70.0 $ & $ 66.4  $ &  2006 Apr 09 & S  &  M105  &      $ 4448 $ & $  30.1$ & $ 29.7  $ &  2003 Sep 01 & I  &  Sagittarius\\
\hline \\
\end{tabular}
\end{minipage}
\label{tab:list}
$^\star$``I'' and ``S'' denote observations performed with ACIS-I and ACIS-S detectors.
\end{table*}  

\section{Sample selection and data reduction}
\subsection{Sample selection}
\label{sec:sample}
The superb angular resolution combined with the low and stable instrumental background of \textit{Chandra} observatory makes the satellite perfectly suitable for the present study. We searched the \textit{Chandra} archive for  observations in the science category ``Normal Galaxies'' and selected a sample of early-type galaxies with point source detection sensitivity better than $ 10^{37} \ \mathrm{erg \ s^{-1}} $. This threshold was chosen to minimize the contribution of unresolved low-mass X-ray binaries (LMXBs), and its particular value is explained later in this paper  (Sect. \ref{sec:xtokvalues}). The sample was further extended to include  the bulge of M31, which has similar stellar population and gas and dust content to elliptical galaxies. To explore young elliptical galaxies    we also added NGC3377 and NGC3585, which would otherwise not pass our selection criteria because of the high point source detection sensitivity.

The sample constructed this way includes  14 galaxies and galaxy bulges, whose main properties are listed in Table \ref{tab:list2}. The point source detection sensitivity refers to the $ 0.5-8 $ keV band, and it was calculated assuming average  spectrum of LMXBs -- a power law with $ \Gamma =1.56 $ \citep{irwin2}.

\subsection{\textit{Chandra}}
We analyzed 70 archival \textit{Chandra} observations, as listed in Table \ref{tab:list}. The total exposure time of the data was $ T_{\mathrm{obs}}  \approx 2.8 $ Ms. For ACIS-I observations we extracted data of the entire ACIS-I array, while for ACIS-S observations we used only the S3 chip. The data reduction was performed with standard CIAO\footnote{http://cxc.harvard.edu/ciao} software package tools (CIAO version 4.1; CALDB version 4.1.3). 

The main steps of the data reduction are similar to those outlined in \citet{bogdan}. First, the flare contaminated time intervals were removed, which decreased the exposure time by $ \lesssim 10 $ per cent. For the point source detection, we used the unfiltered data, because longer exposure time outweighs higher background periods. We combined the available data for each galaxy by projecting them into the coordinate system of the observation with the longest exposure time. We ran the CIAO \texttt{wavdetect} tool on the merged data in the $ 0.5-8 $ keV energy range, and applied the same parameters as in \citet{bogdan}. This  results in relatively large source cells, in order to minimize the contribution of residual point source counts to the unresolved emission. We find that $ \gtrsim 98 $ per cent of the source counts lie within the obtained regions. 
Because of the large cell size, some of the source cells may overlap in the central regions of some of the galaxies from our sample. However, this is not a problem, because the goal of this analysis is the unresolved emission, rather than point sources.
The source list was used to mask out point sources in the analysis of the unresolved emission. 

The background subtraction plays a crucial role in studying the extended X-ray emission. In all galaxies, except for M31, we used a combination of several regions away from the galaxies to estimate the sky and instrumental background components. This technique cannot be applied in M31 since its angular size is significantly greater than the field of view of the detectors, therefore we followed the procedure described in \citet{bogdan}. We constructed exposure maps by assuming a power-law model with a slope of $\Gamma=2$.

\subsection{Near-infrared data}
To trace the stellar light distribution, we used the K-band data of 2MASS Large Galaxy Atlas \citep{jarrett}. Because of the large angular size of M31, the background is somewhat oversubtracted on its near-infrared images (T. Jarrett, private communication), so that the outer bulge and disk of the galaxy appears to be too faint. To avoid the uncertainties caused by the improper background subtraction, we used the Infrared Array Camera (IRAC) onboard \textit{Spitzer Space Telescope} (\textit{SST}) to trace the stellar light in M31. To facilitate the comparison of results with other galaxies in our sample, we converted the observed near-infrared \textit{Spitzer} luminosities to K-band values. The scale between the $ 3.6 \ \mathrm{\mu m} $ IRAC and 2MASS K-band data was determined in the center of M31, where the background level is negligible. The obtained conversion factor between the pixel values is $ C_{K}/C_{3.6 \ \mathrm{\mu m}} \approx 10.4 $. 

\begin{figure*}[!ht]
\hbox{
\includegraphics[width=6.35cm]{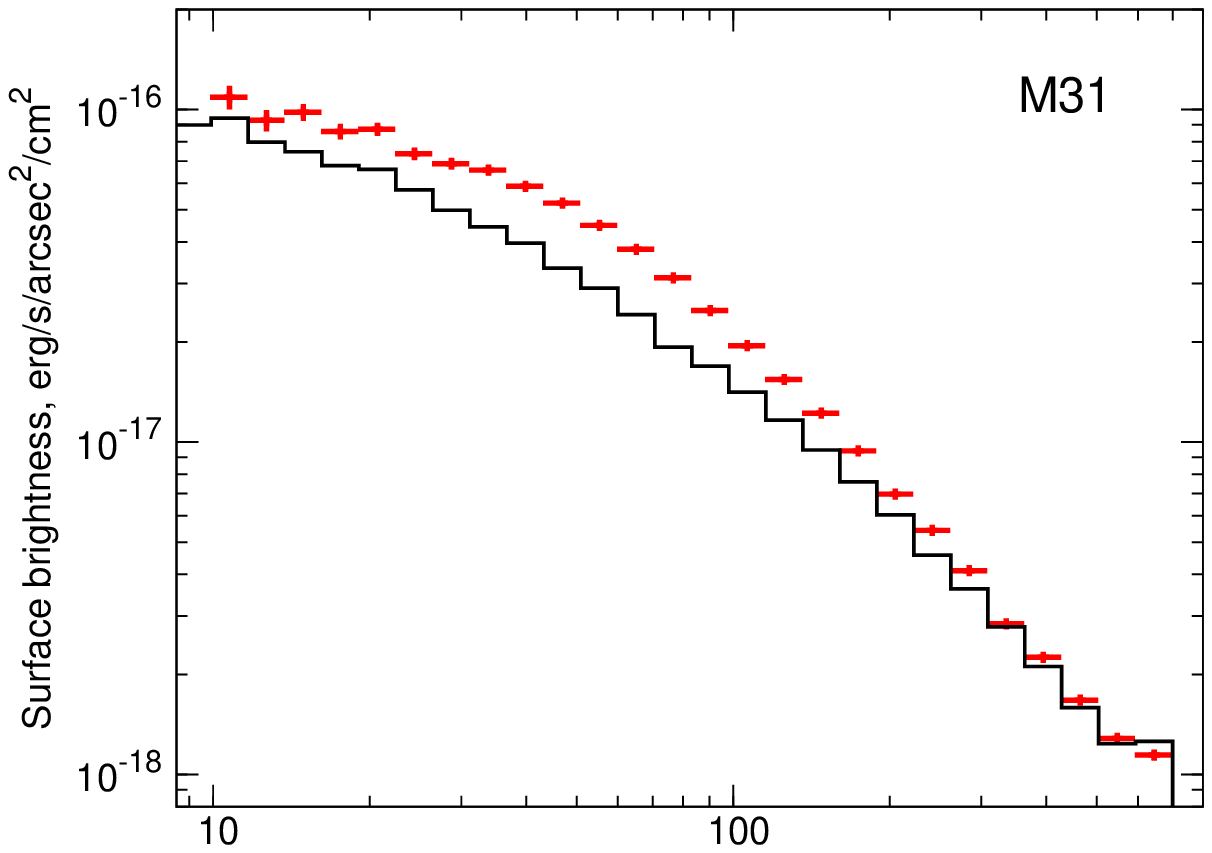}
\hspace{-0.45cm}
\includegraphics[width=6.35cm]{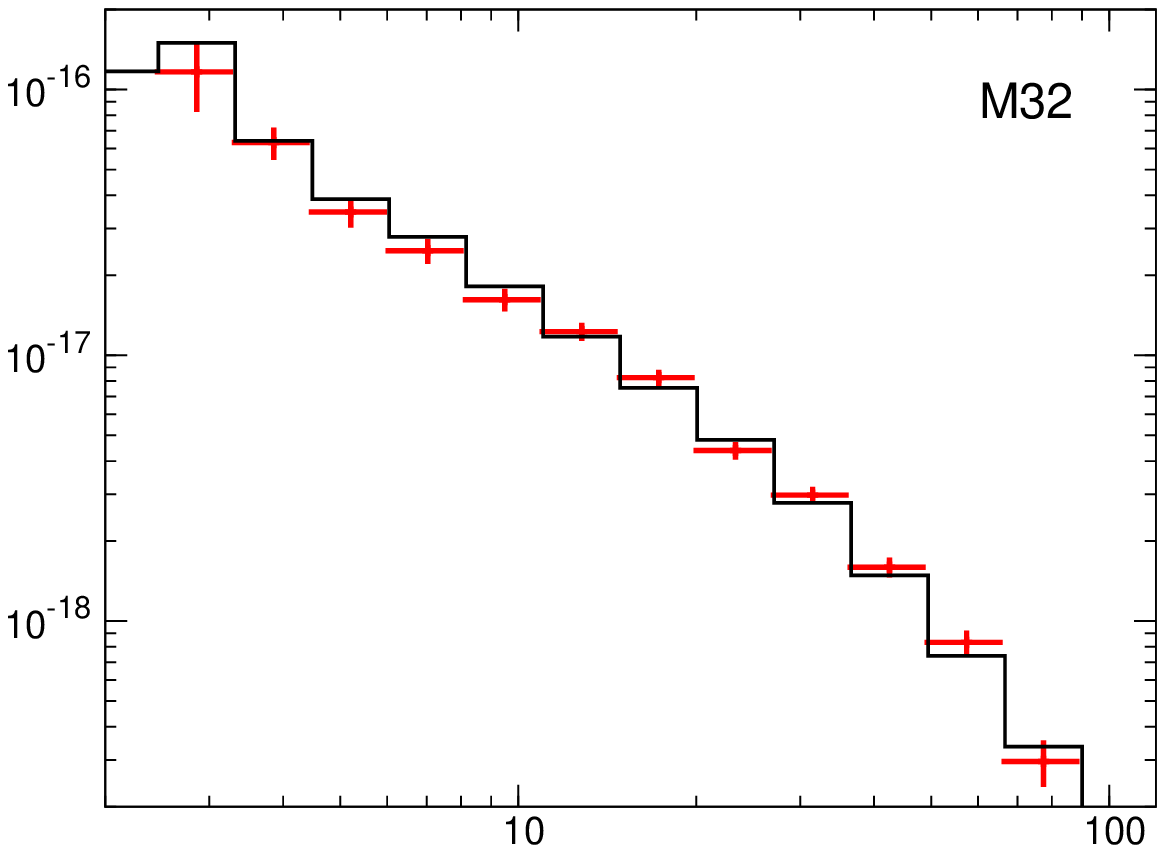}
\hspace{-0.45cm}
\includegraphics[width=6.35cm]{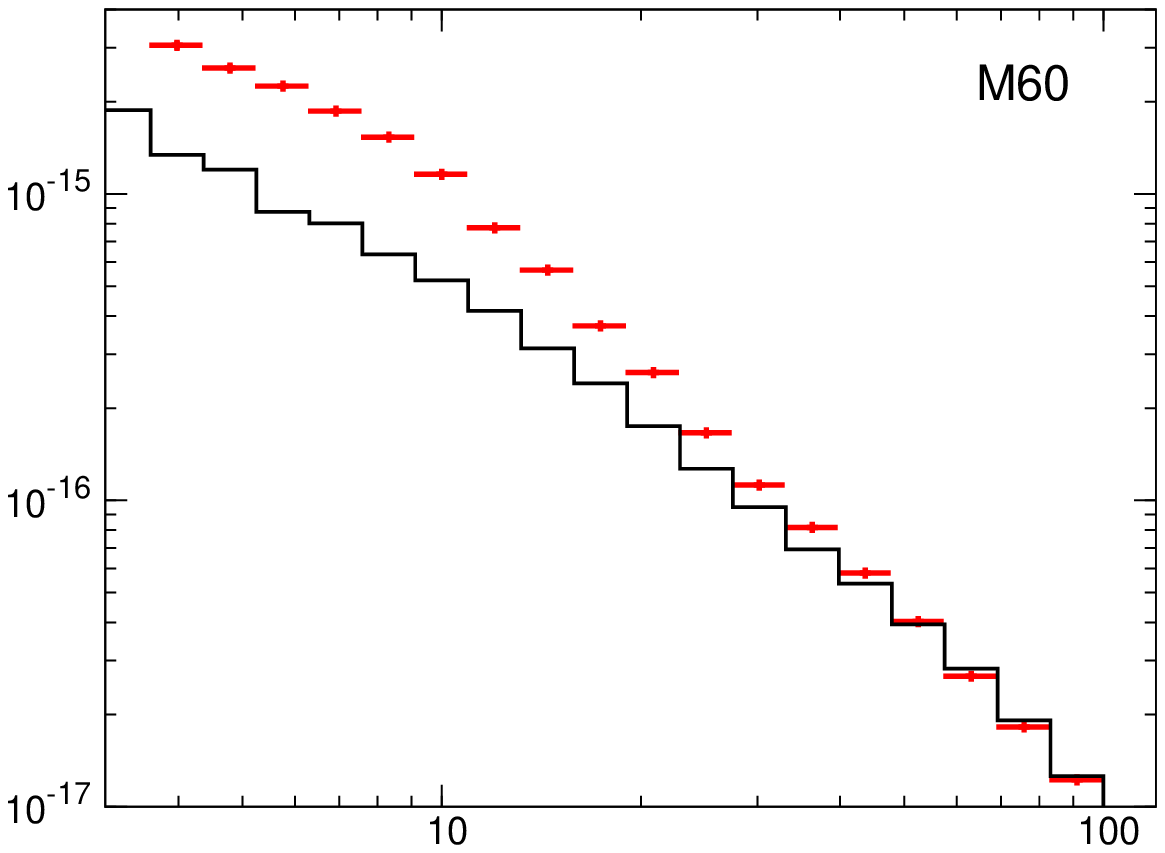}
}
\hbox{
\includegraphics[width=6.35cm]{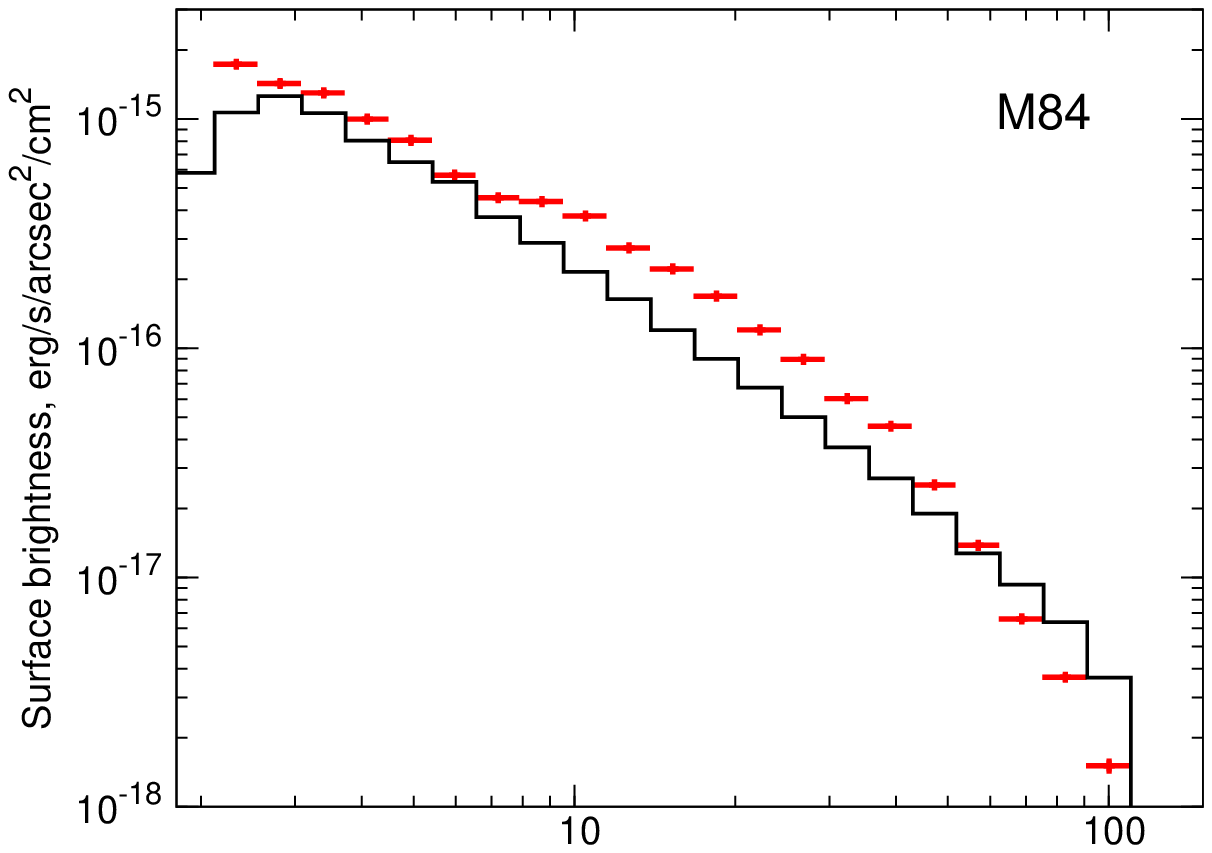}
\hspace{-0.45cm}
\includegraphics[width=6.35cm]{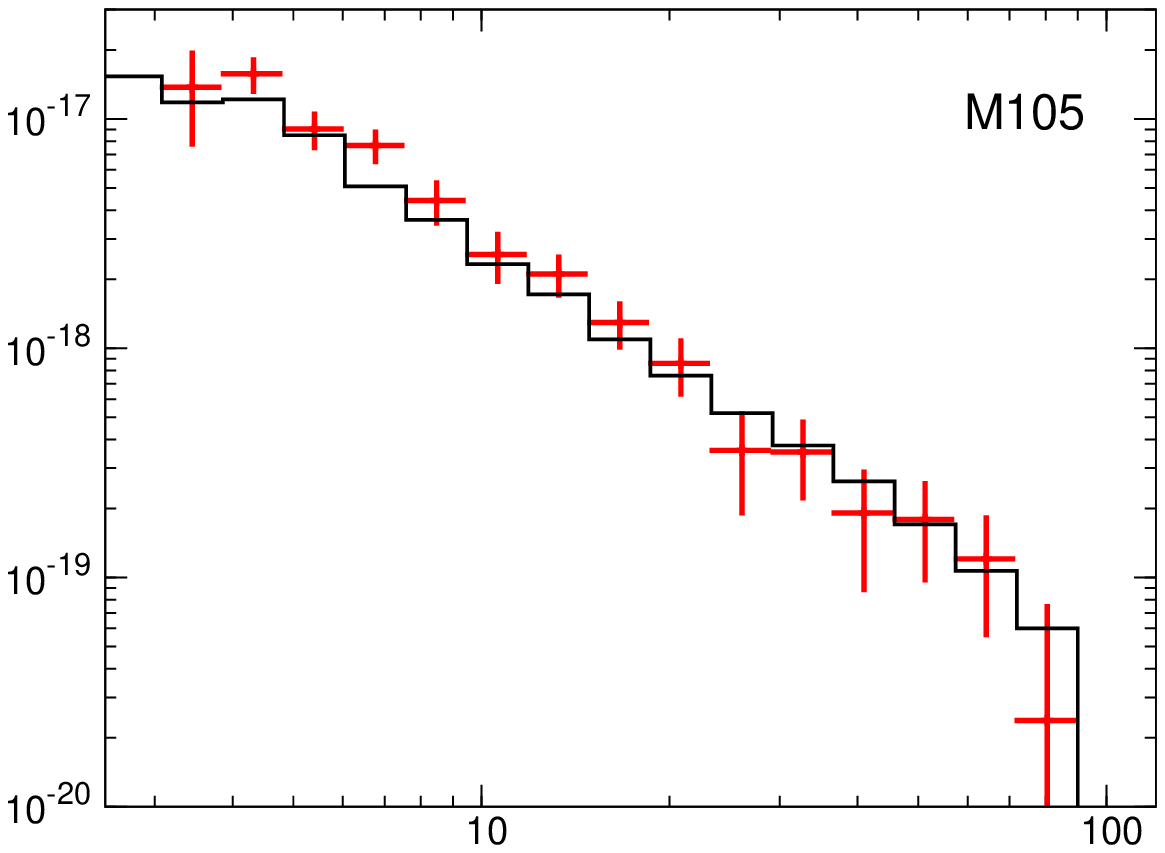}
\hspace{-0.45cm}
\includegraphics[width=6.35cm]{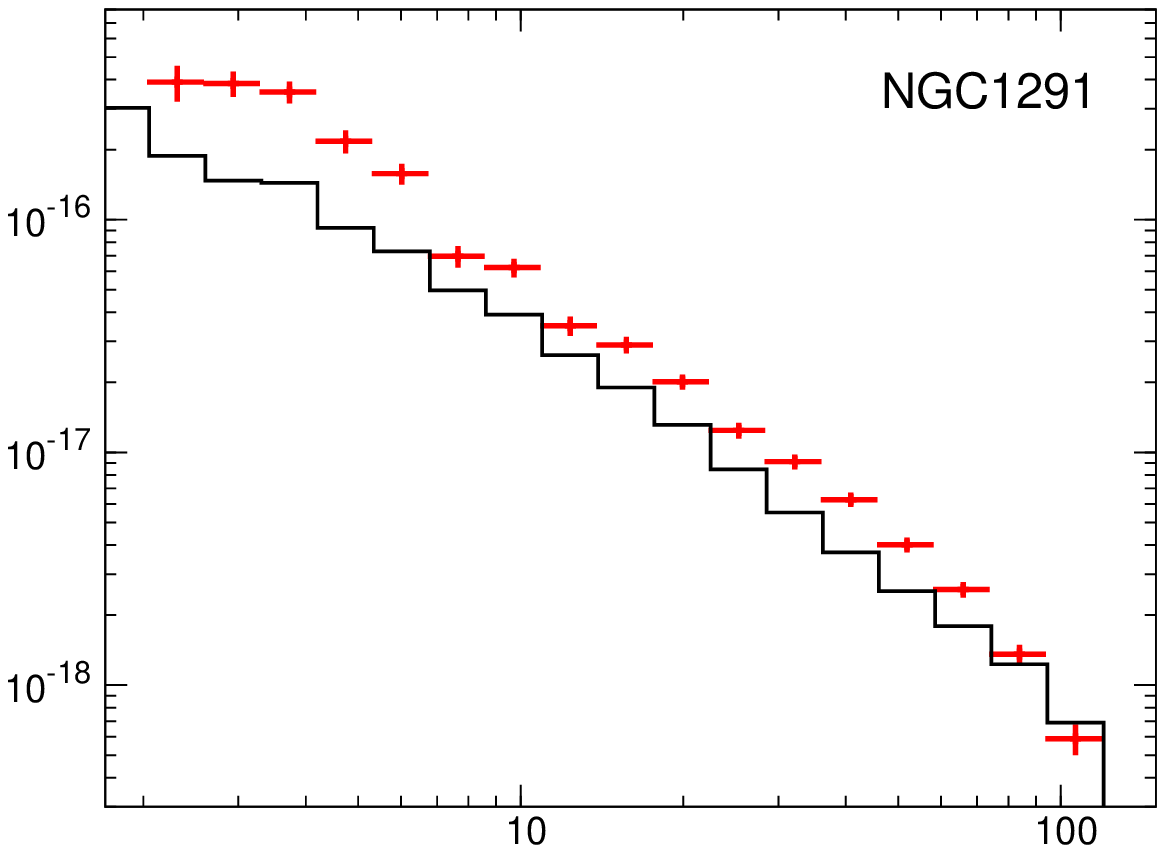}
}
\hbox{
\includegraphics[width=6.35cm]{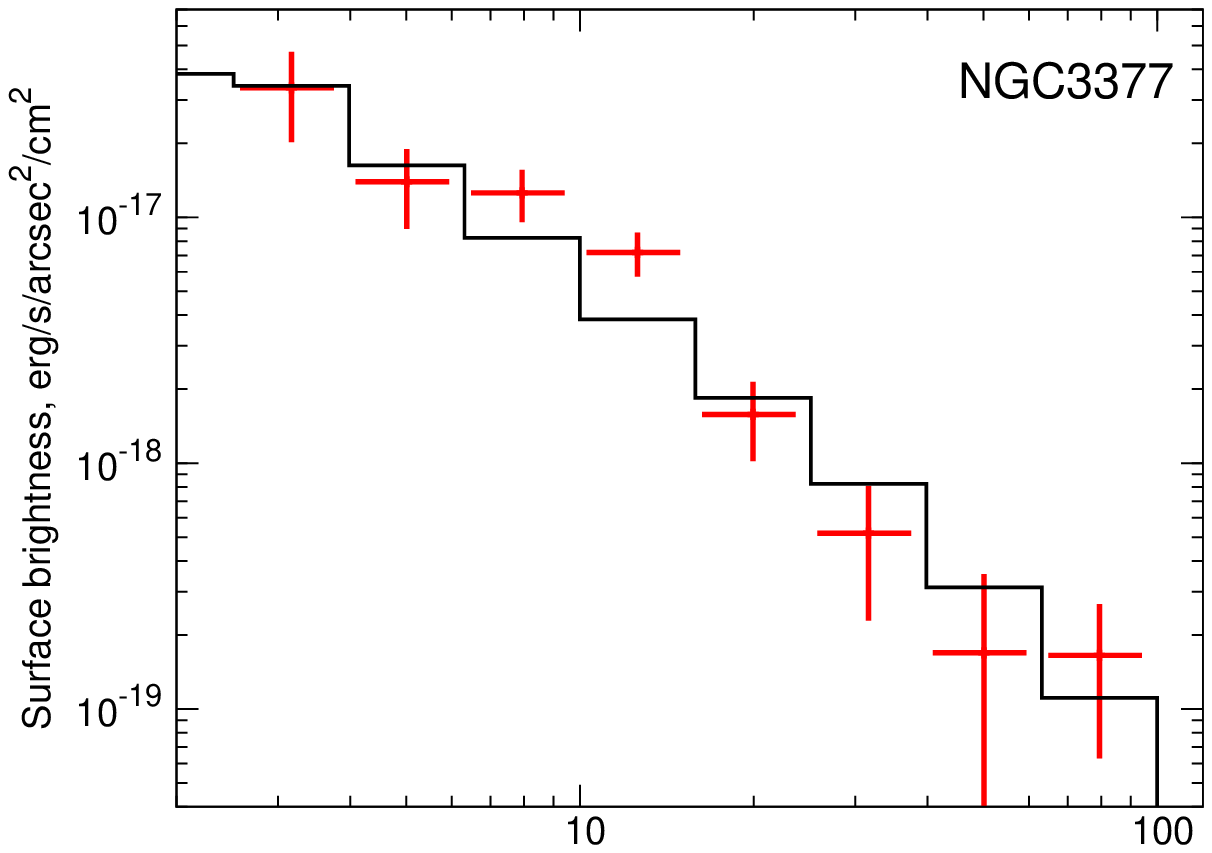}
\hspace{-0.45cm}
\includegraphics[width=6.35cm]{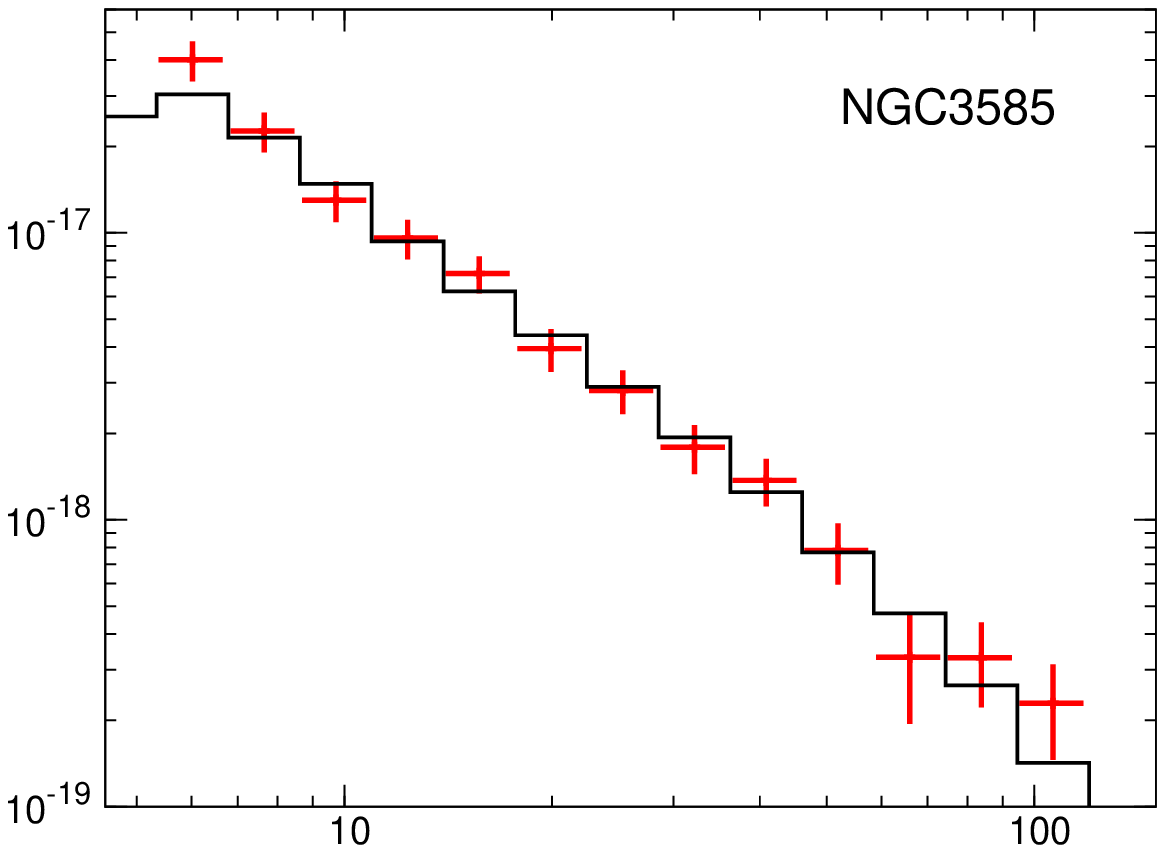}
\hspace{-0.45cm}
\includegraphics[width=6.35cm]{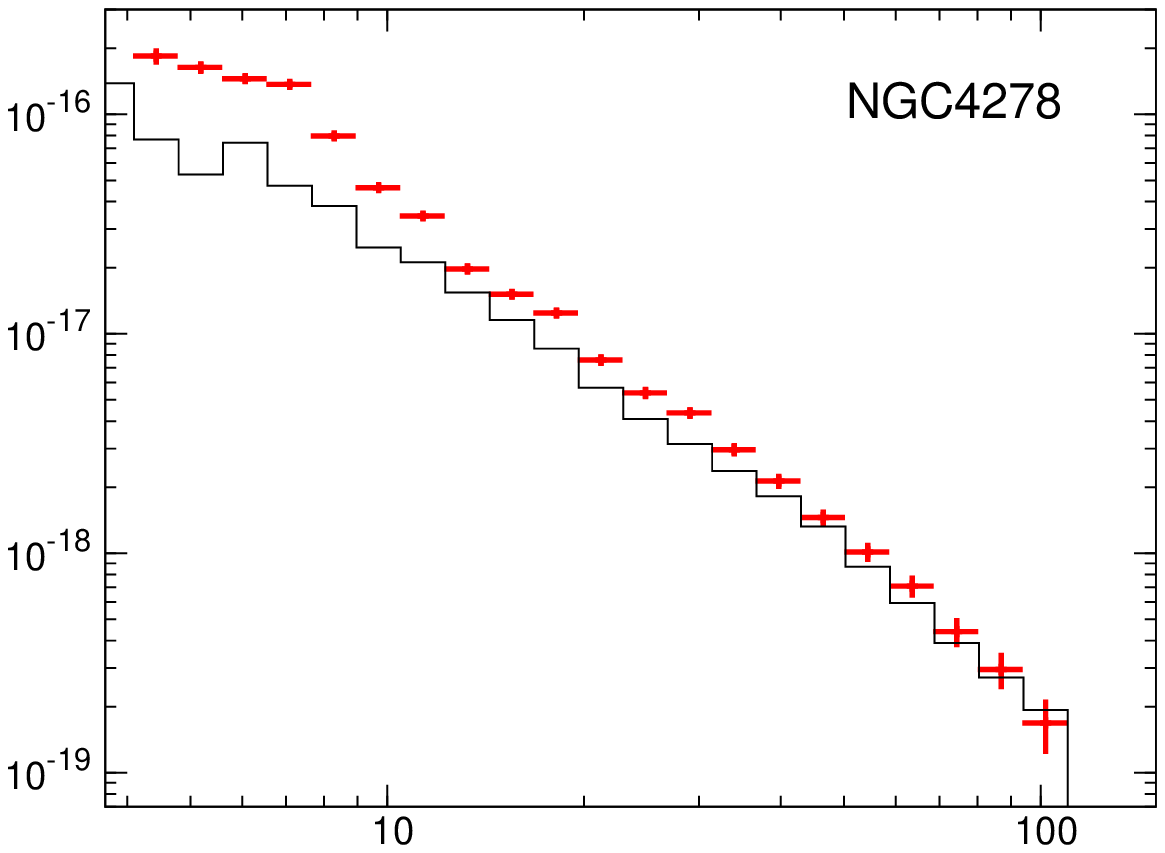}
}
\hbox{
\includegraphics[width=6.35cm]{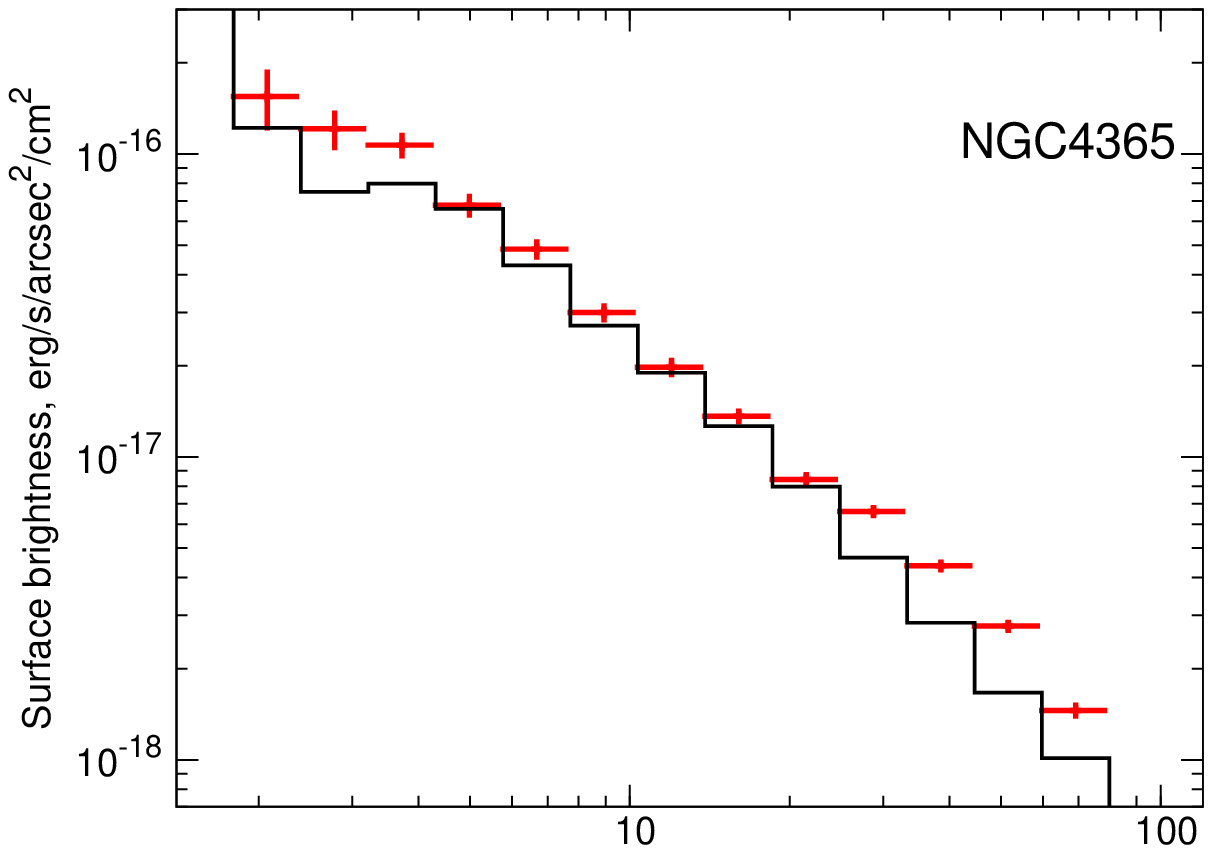}
\hspace{-0.45cm}
\includegraphics[width=6.35cm]{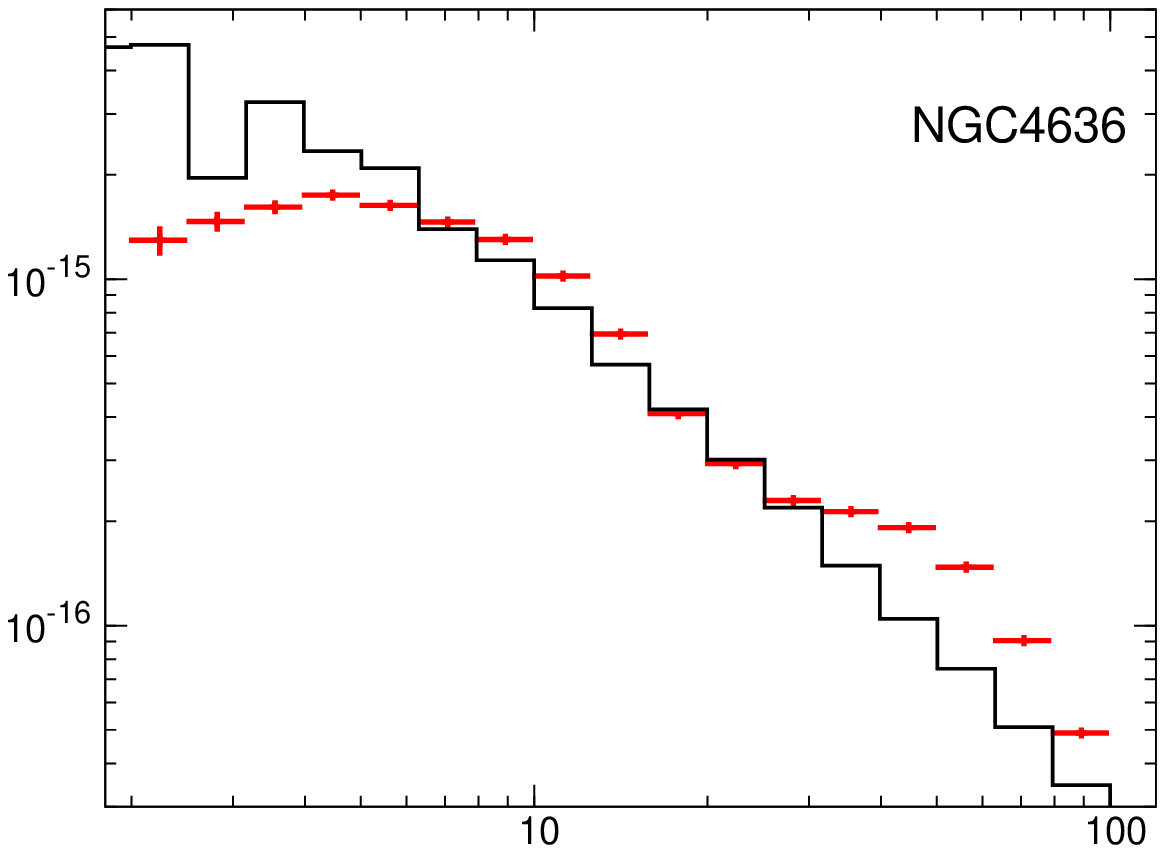}
\hspace{-0.45cm}
\includegraphics[width=6.35cm]{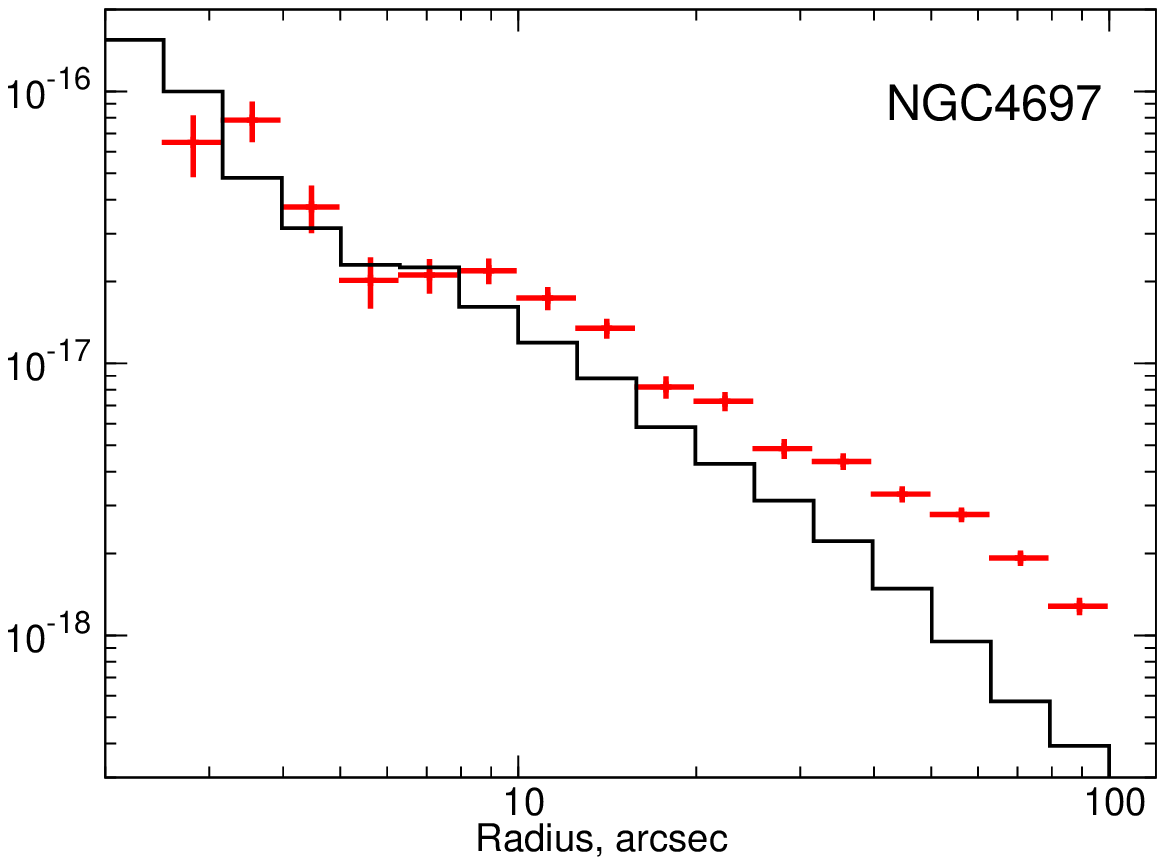}
}
\hbox{
\includegraphics[width=6.35cm]{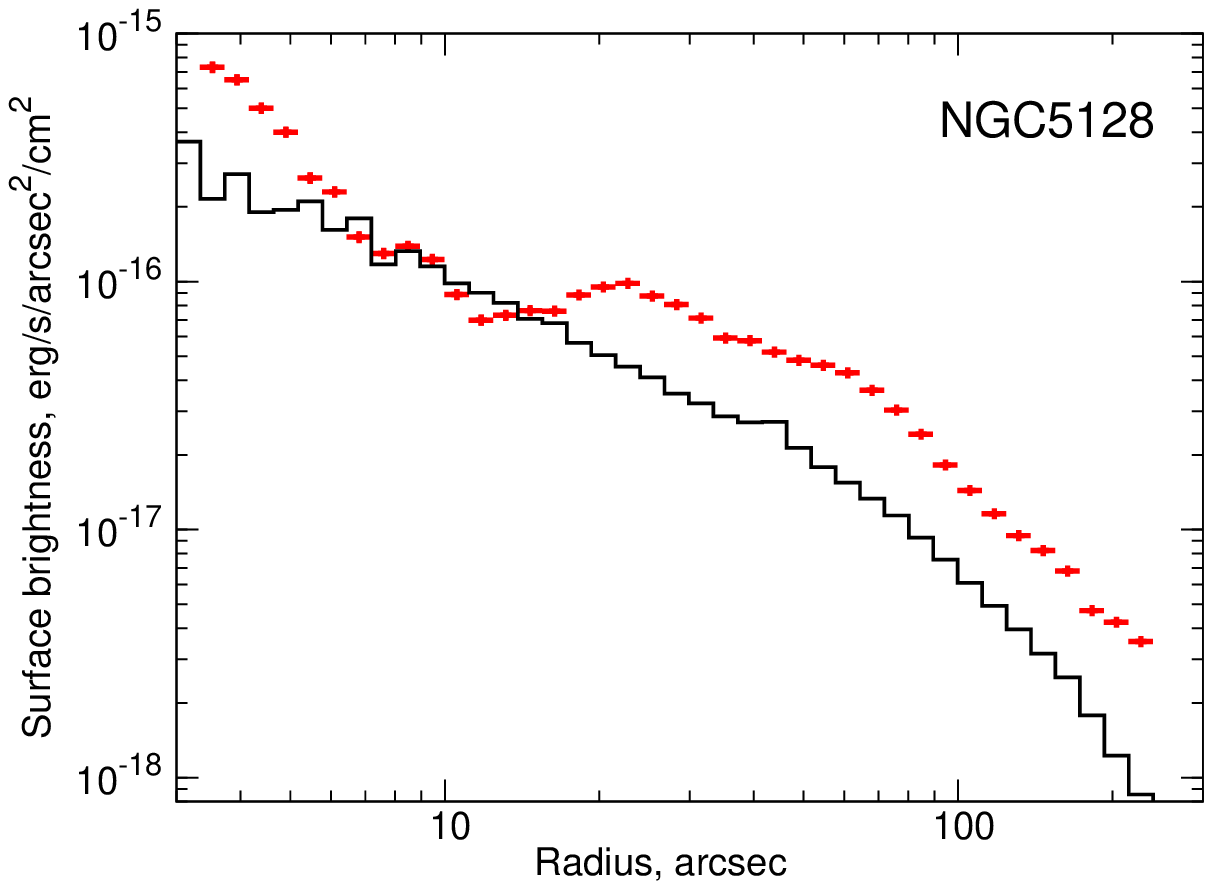}
\hspace{-0.45cm}
\includegraphics[width=6.35cm]{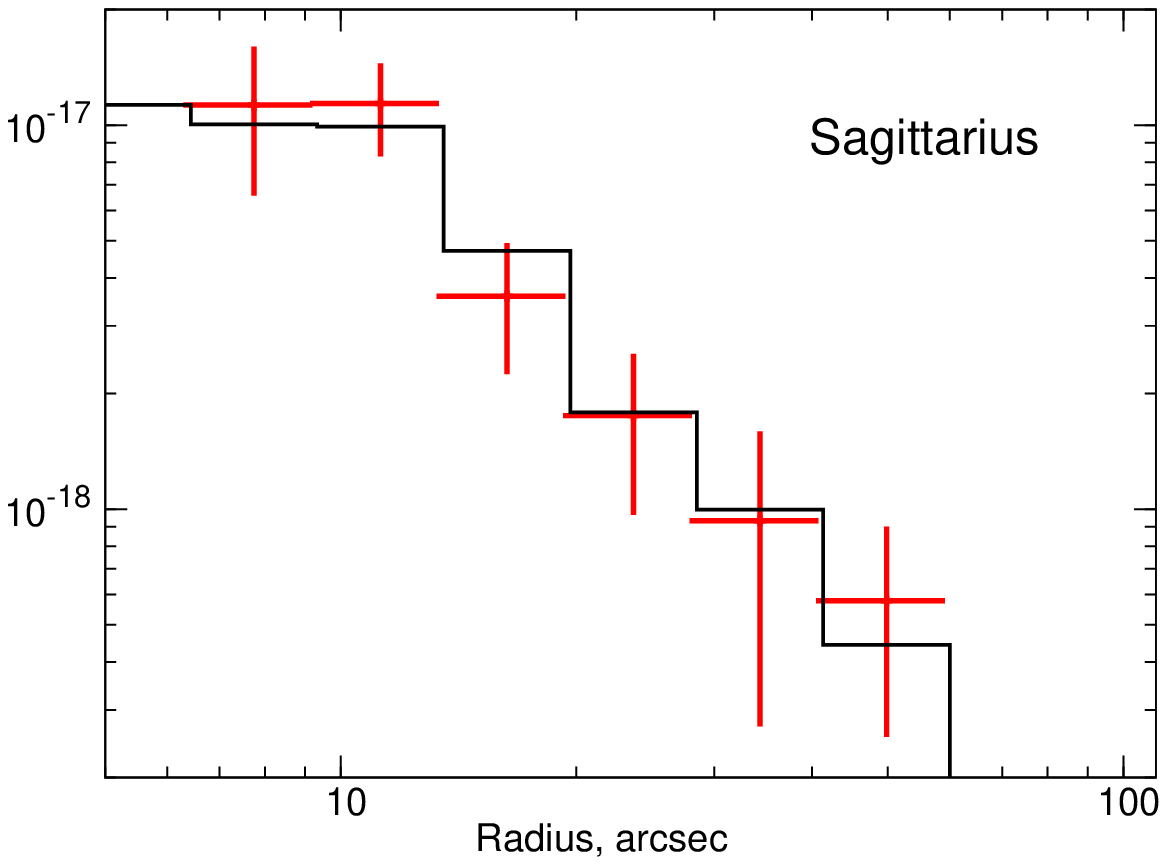}
}
\caption{Surface brightness profiles of the 14 galaxies in our sample in the $ 0.5-2 $ keV energy range. All background components were subtracted. Crosses (red) show the distribution of unresolved X-ray emission based on \textit{Chandra} data, the solid line (black) shows the (arbitrarily) re-normalized near-infrared brightness.}  
\label{fig:profiles}
\end{figure*}

\section{Identifying gas-poor galaxies}
Truly diffuse emission is present in many luminous elliptical galaxies, that originates from warm ionized gas. Its luminosity may significantly exceed the emission from unresolved compact X-ray sources. Therefore our aim is to identify the emission of warm ionized gas, and, if possible, to separate it from unresolved faint compact sources. To reveal the presence and map the distribution of the ISM we constructed radial surface brightness profiles of the unresolved X-ray emission and studied its spectra. 

In Fig. \ref{fig:profiles} the distribution of X-ray surface brightness is shown for the studied galaxies. The profiles were extracted in the $ 0.5-2 $ keV energy band using circular annuli, centered on the center of the galaxy. The data was corrected for vignetting, and all background components were subtracted. The contribution of resolved compact sources was removed as described in the previous section. 

\begin{figure*}[!ht]
\hbox{
\includegraphics[width=6.35cm]{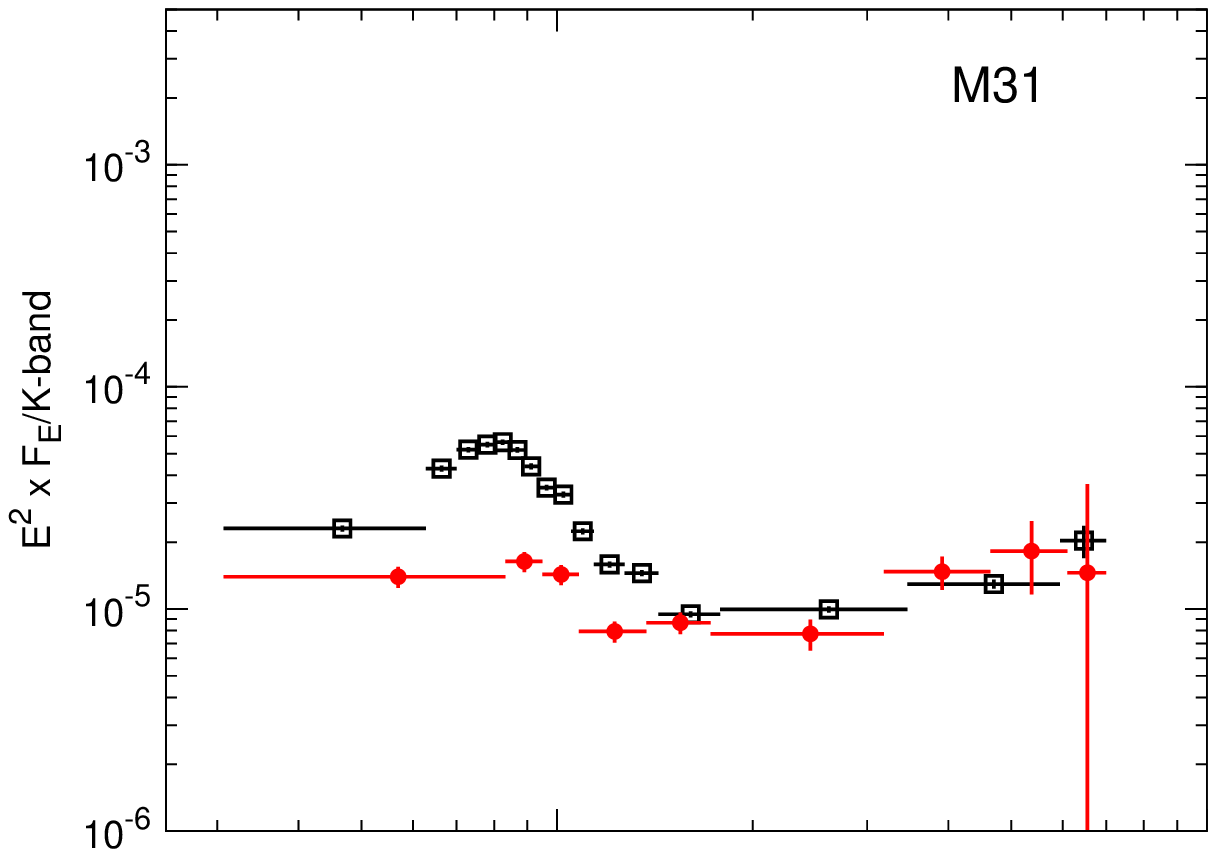}
\hspace{-0.45cm}
\includegraphics[width=6.35cm]{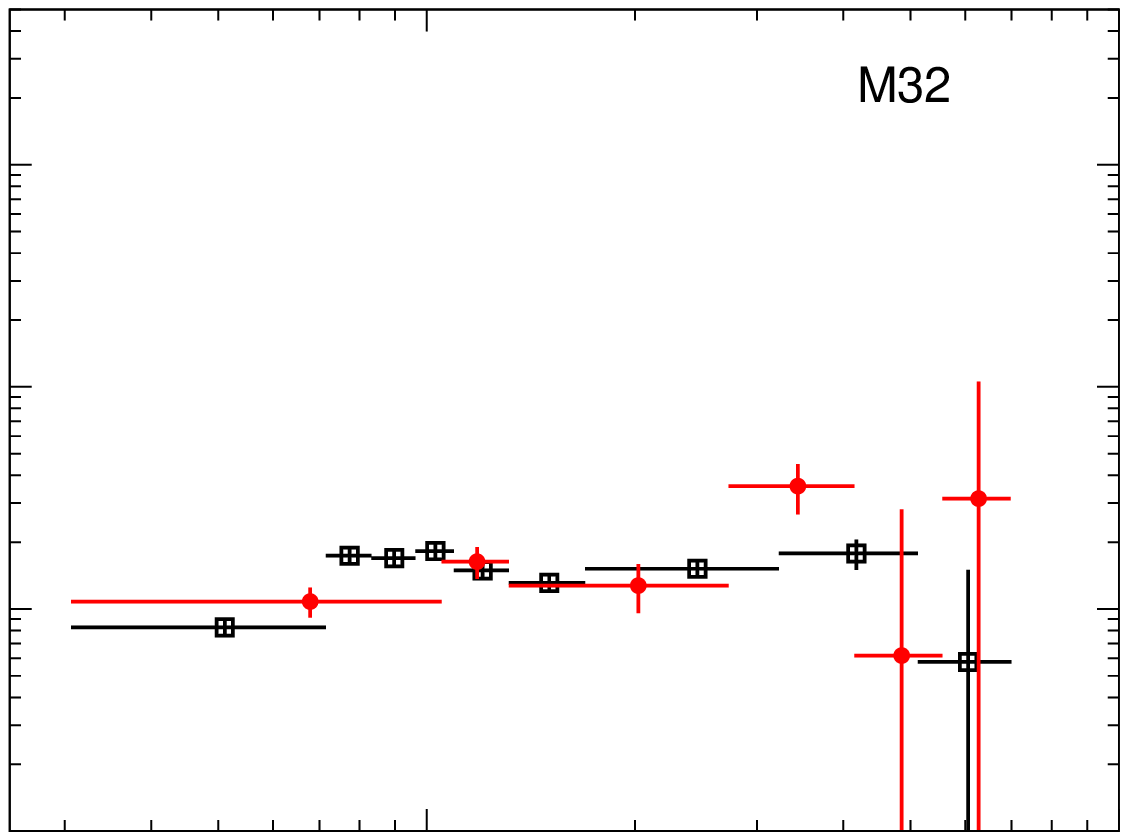}
\hspace{-0.45cm}
\includegraphics[width=6.35cm]{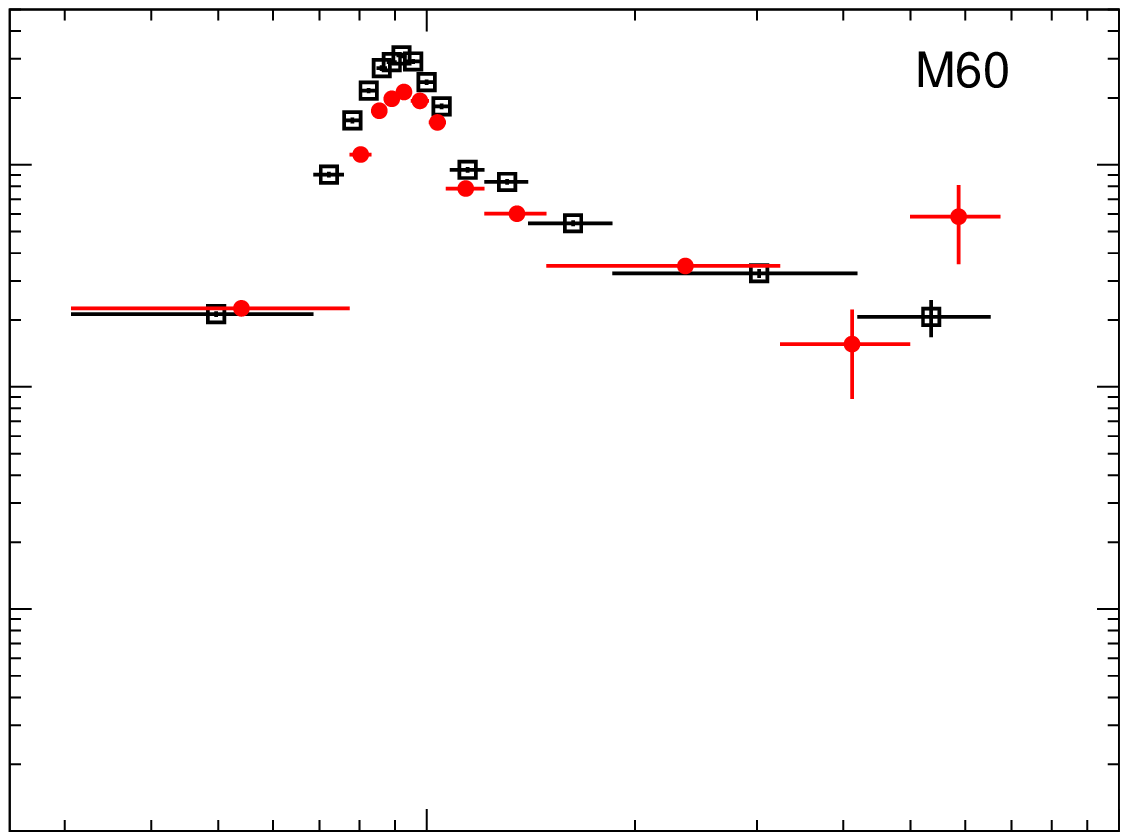}
}
\hbox{
\includegraphics[width=6.35cm]{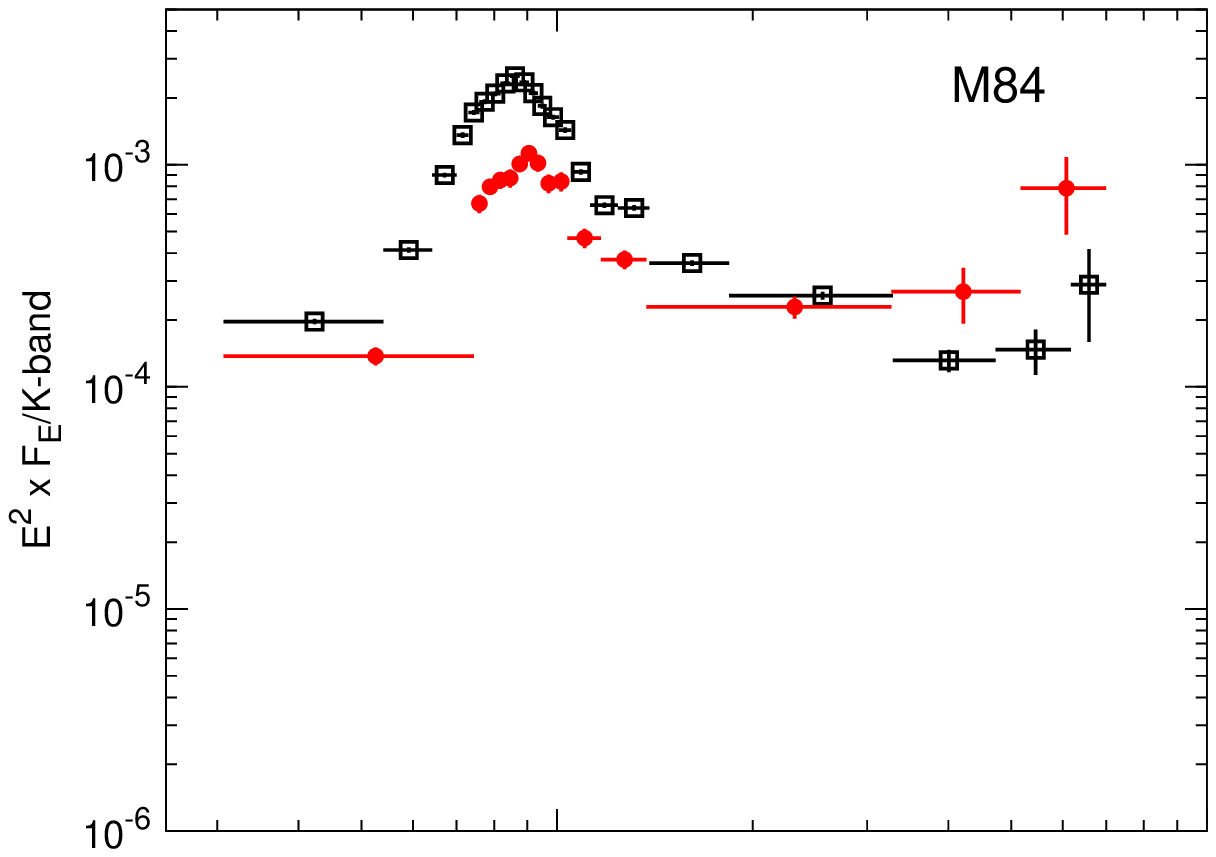}
\hspace{-0.45cm}
\includegraphics[width=6.35cm]{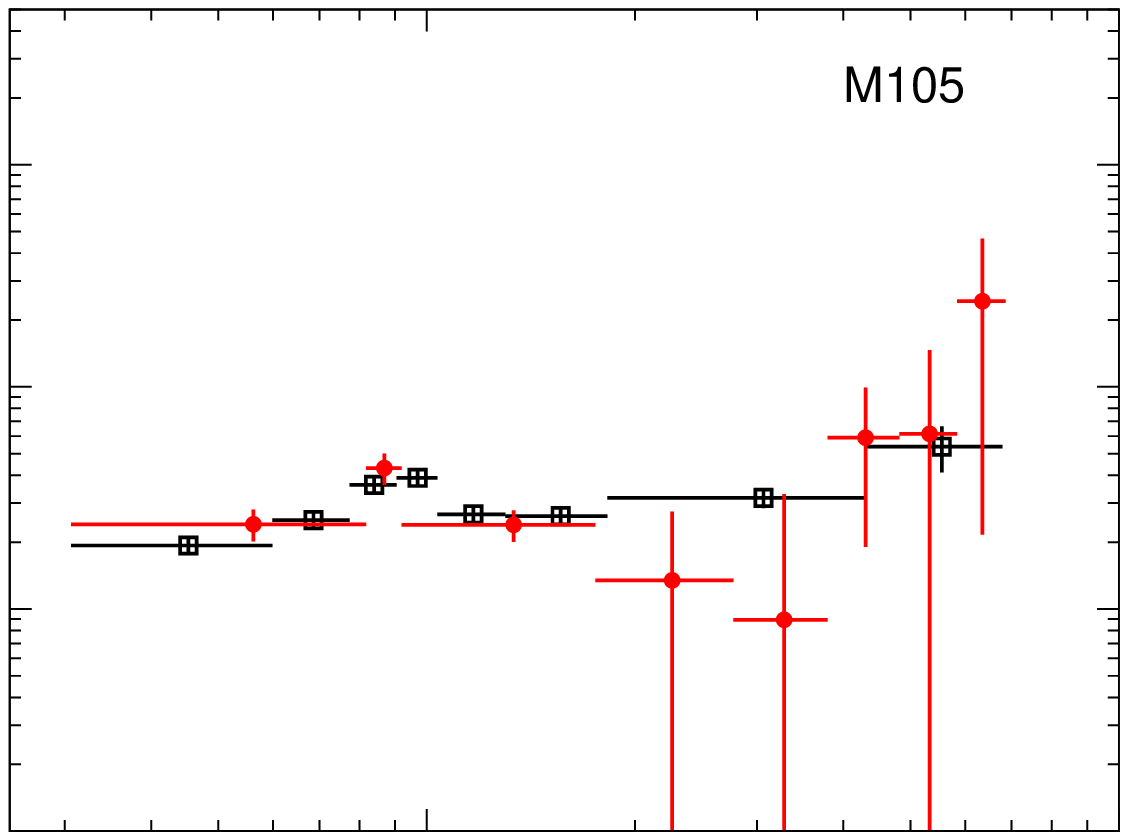}
\hspace{-0.45cm}
\includegraphics[width=6.35cm]{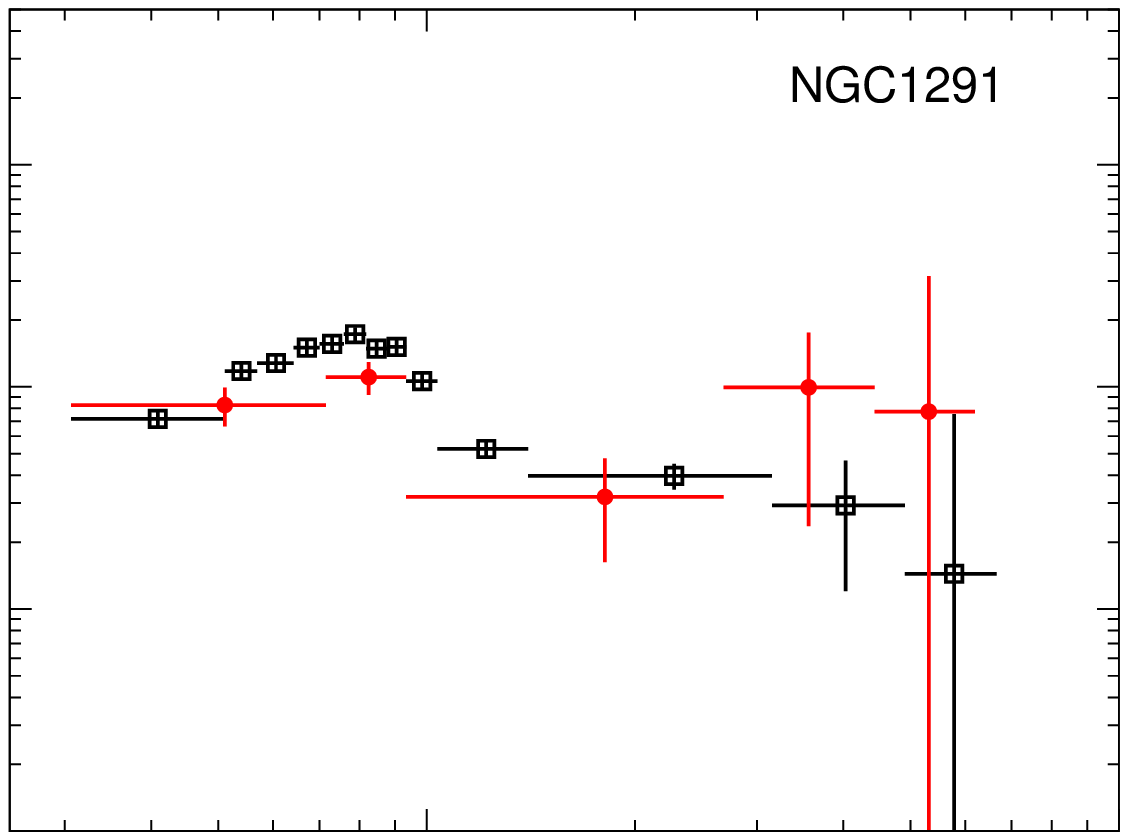}
}
\hbox{
\includegraphics[width=6.35cm]{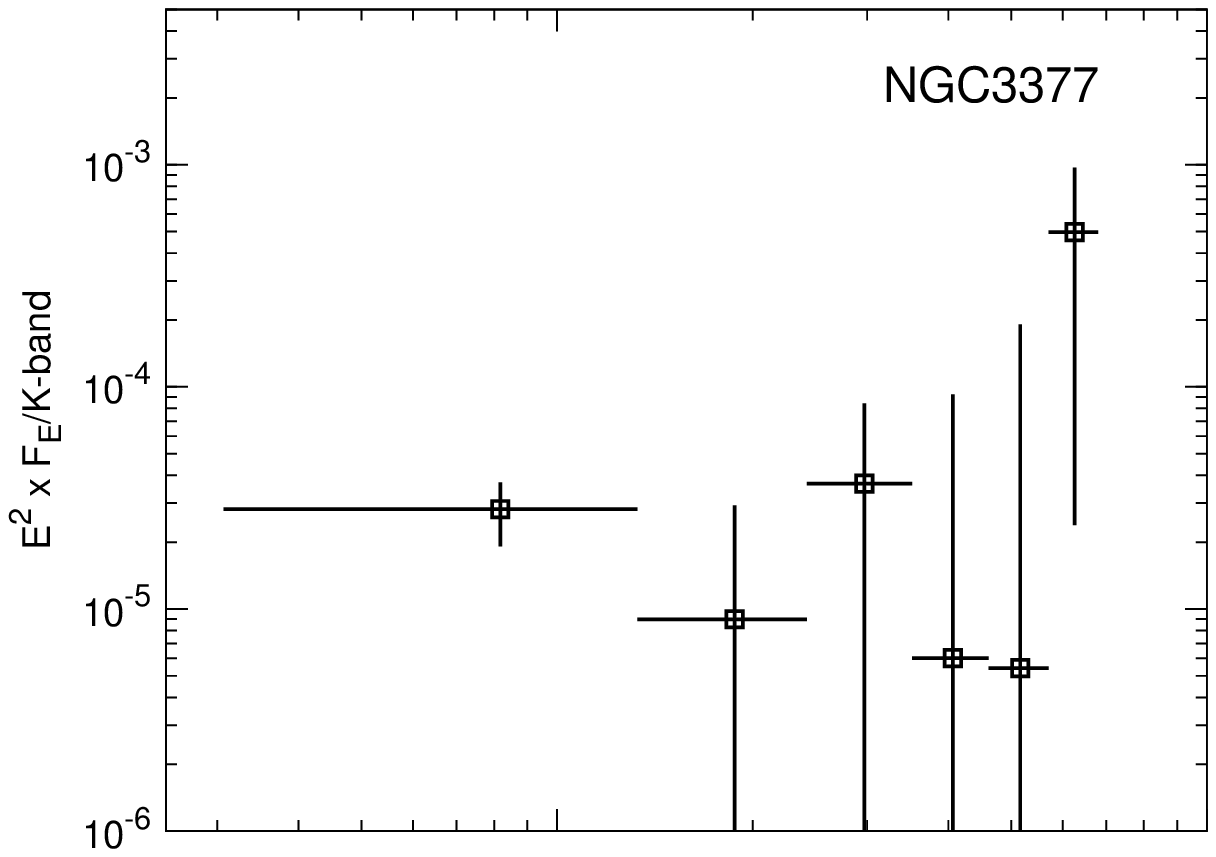}
\hspace{-0.45cm}
\includegraphics[width=6.35cm]{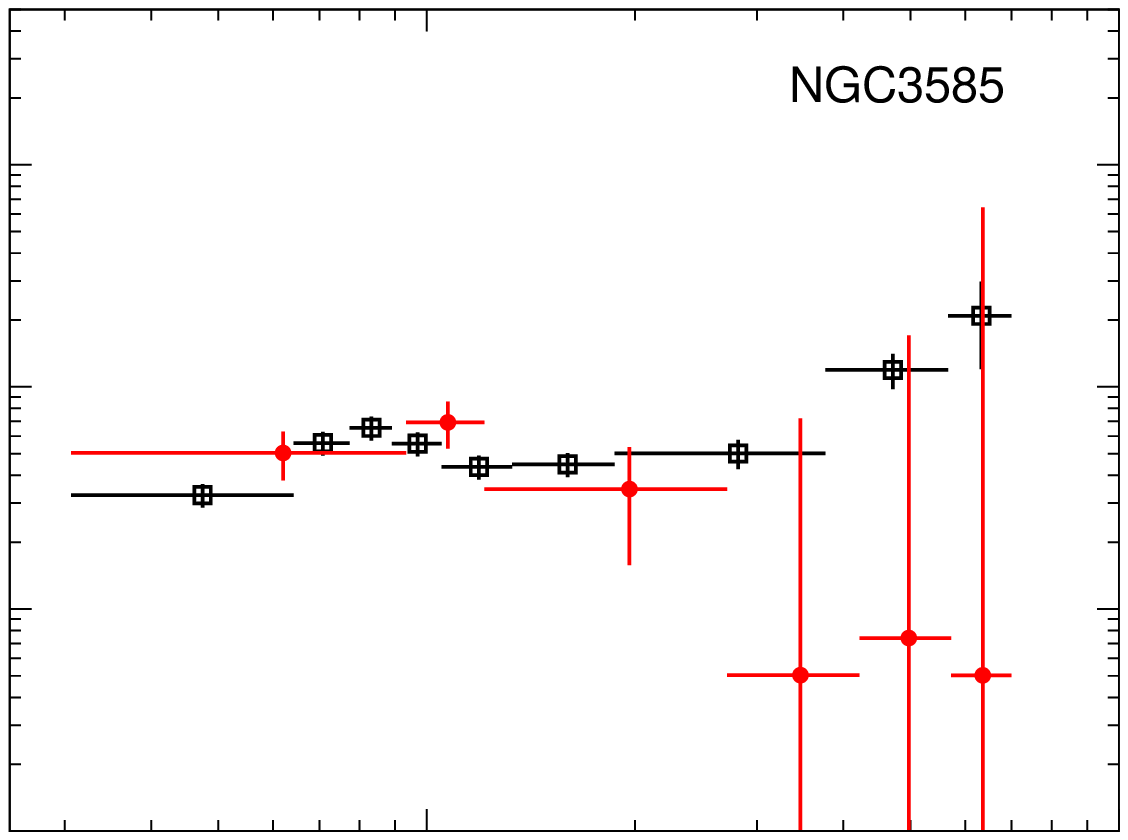}
\hspace{-0.45cm}
\includegraphics[width=6.35cm]{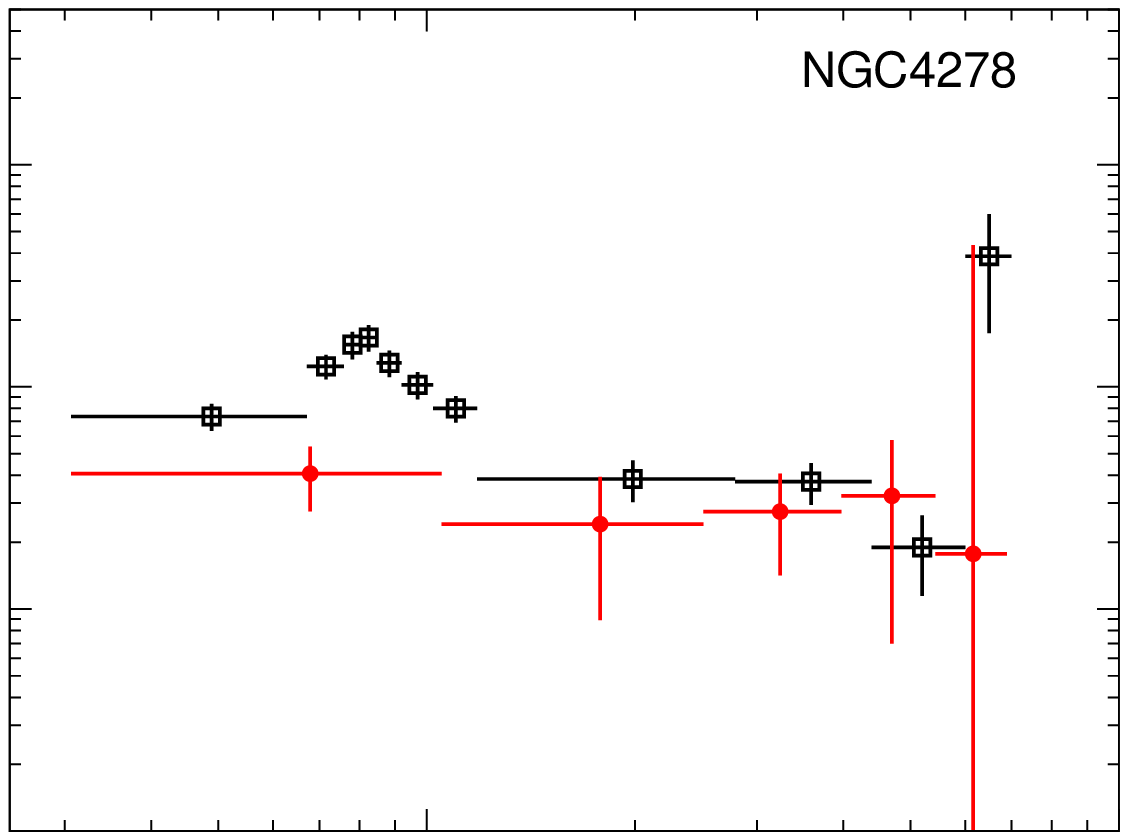}
}
\hbox{
\includegraphics[width=6.35cm]{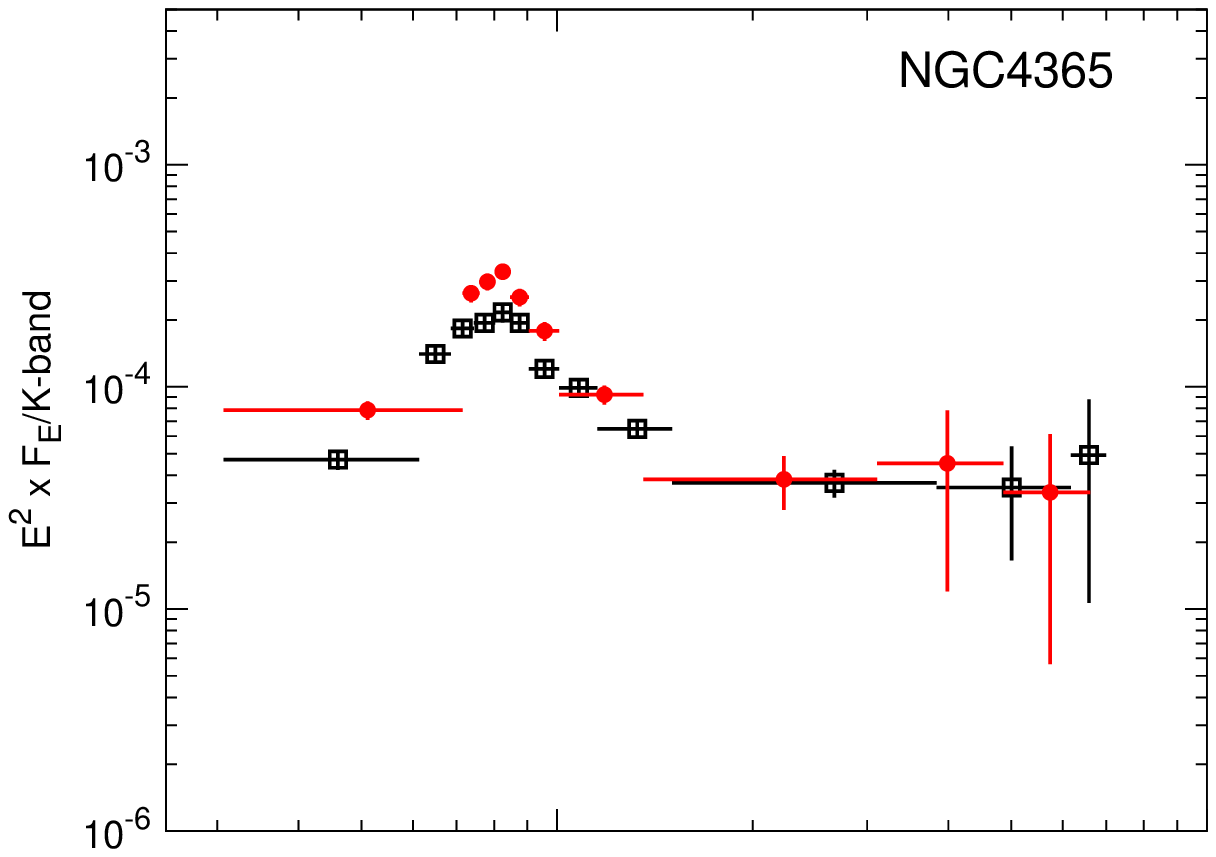}
\hspace{-0.45cm}
\includegraphics[width=6.35cm]{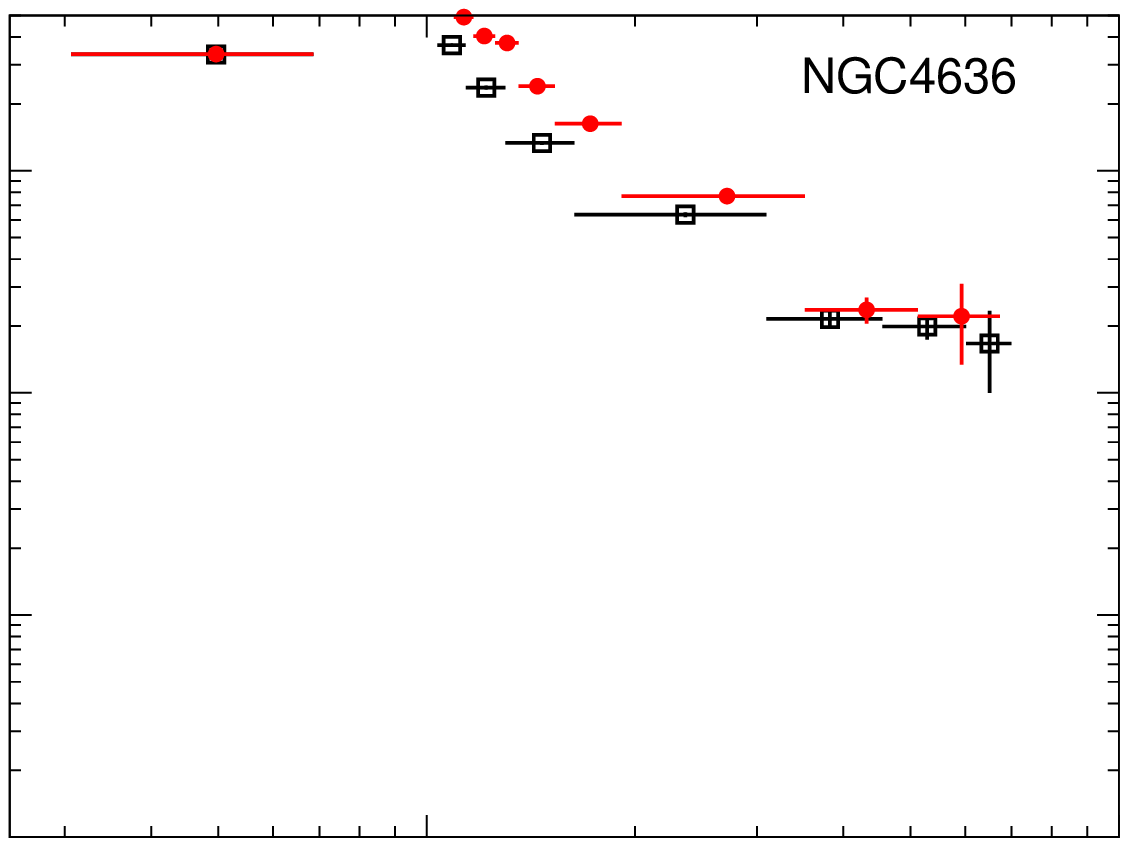}
\hspace{-0.45cm}
\includegraphics[width=6.35cm]{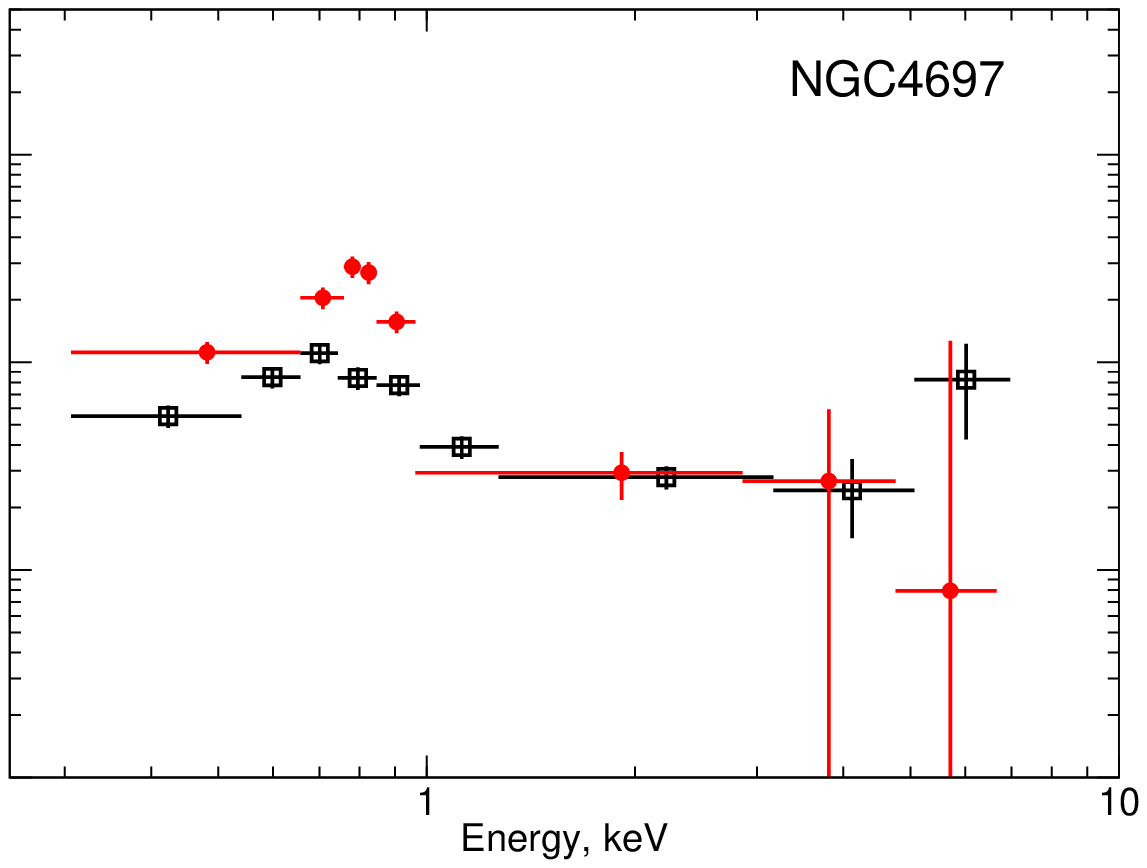}
}
\hbox{
\includegraphics[width=6.35cm]{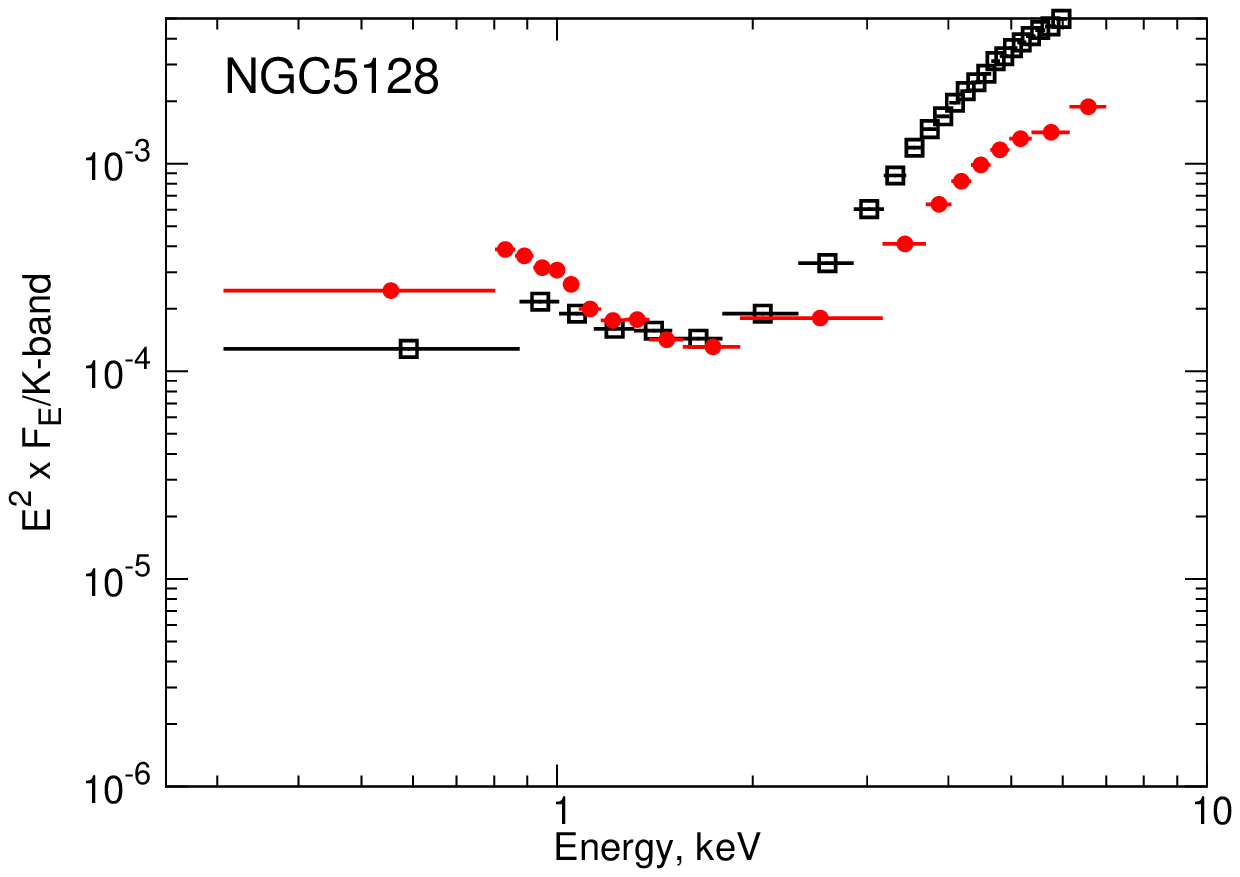}
\hspace{-0.45cm}
\includegraphics[width=6.35cm]{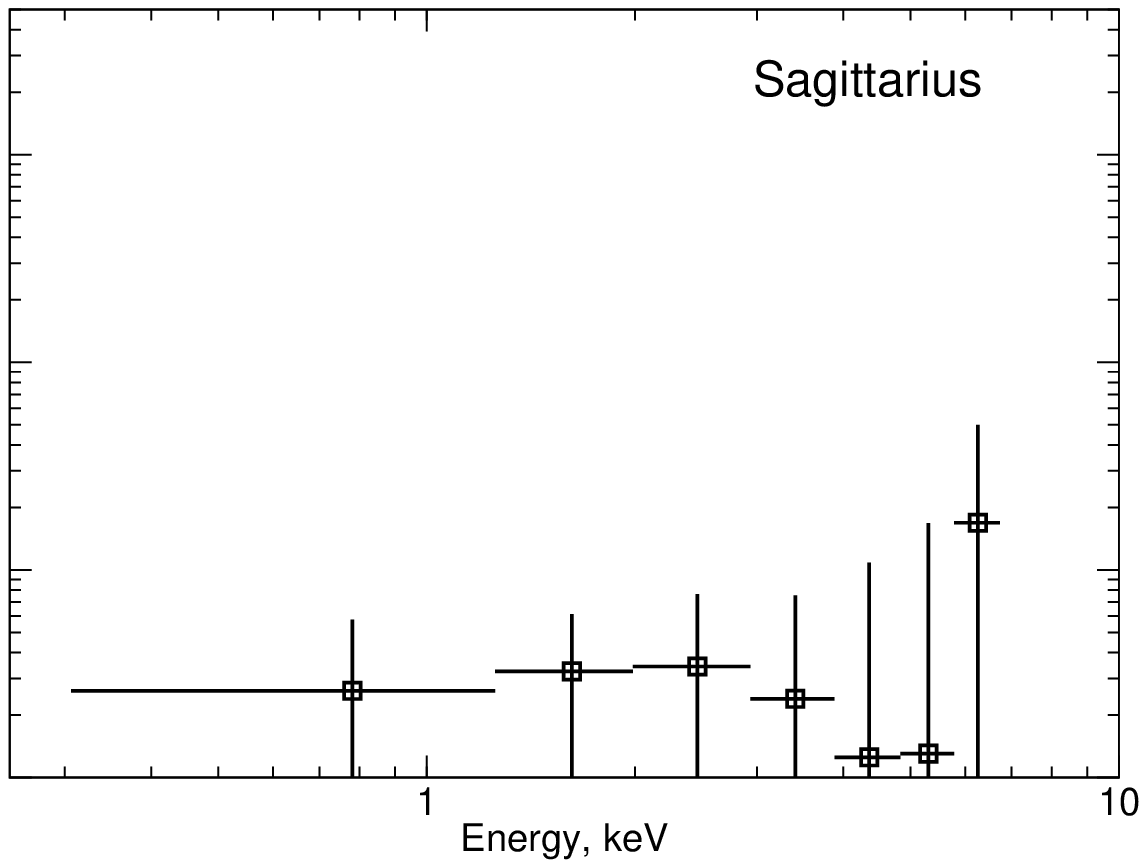}
}
\caption{Energy spectra of the 14 galaxies in our sample. All spectra were normalized to the same level of near-infrared brightness and to an assumed distance of 10 Mpc, and background is subtracted. The scales on the \textit{x}- and \textit{y}-axes are the same in all panels. For each galaxy, hollow squares  (black) show the spectrum of the inner region, while filled circles (red) represent the outer region spectrum. In the case of NGC3377 and Sagittarius  spectra of the entire galaxy are shown because of the relatively low number of counts.}  
\label{fig:spectra}
\end{figure*}

\begin{table*}[!ht]
\caption{Results of spectral fits of unresolved emission.}
\renewcommand{\arraystretch}{1.4}
\begin{minipage}{18cm}
\centering
\begin{tabular}{c c c c c c c c c}
\hline 
Name & $kT_{MKL}$ & $\Gamma$ & $L_{MKL}$  & $ L_{PL} $ & $ L_{LMXB} $  & $ L_{MKL}/L_{PL}$ & $ L_{MKL}/(L_{PL}-L_{LMXB})$ & $ \chi^2 $/d.o.f   \\
& (keV) &  & ($ \mathrm{erg \ s^{-1}} $) & ($ \mathrm{erg \ s^{-1}} $)  & ($ \mathrm{erg \ s^{-1}} $) & & &\\ 
& (1)   &(2)&            (3)             &      (4)                     &        (5)                 &(6)&(7)&(8) \\
\hline 
M31 bulge  & $ 0.32 \pm 0.01 $ &  $ 1.82 \pm 0.04 $ & $ 1.8 \cdot 10^{38} $ & $ 2.5 \cdot 10^{38} $  &                   --  & $ 0.72 $ &  $ 0.72 $ & $  473/223 $   \\     
M32        & $ 0.60 \pm 0.13 $ &  $ 1.65 \pm 0.15 $ & $ 1.1 \cdot 10^{36} $ & $ 5.3 \cdot 10^{36} $  &                   --  & $ 0.21 $ &  $ 0.21 $ & $   76/70  $   \\     
M60        & $ 0.80 \pm 0.01 $ &  $ 1.70 \pm 0.11 $ & $ 6.2 \cdot 10^{40} $ & $ 1.5 \cdot 10^{40} $  & $ 2.0 \cdot 10^{39} $ & $ 4.13 $ &  $ 4.77 $ & $ 1015/100 $   \\     
M84        & $ 0.65 \pm 0.01 $ &  $ 1.45 \pm 0.12 $ & $ 3.0 \cdot 10^{40} $ & $ 1.1 \cdot 10^{40} $  & $ 1.5 \cdot 10^{39} $ & $ 2.73 $ &  $ 3.16 $ & $  385/122 $   \\     
M105       & $ 0.64 \pm 0.05 $ &  $ 2.26 \pm 0.19 $ & $ 4.7 \cdot 10^{37} $ & $ 1.8 \cdot 10^{38} $  & $ 3.8 \cdot 10^{37} $ & $ 0.26 $ &  $ 0.33 $ & $   74/73  $   \\     
NGC1291    & $ 0.31 \pm 0.03 $ &  $ 2.64 \pm 0.23 $ & $ 6.3 \cdot 10^{38} $ & $ 9.9 \cdot 10^{38} $  & $ 3.2 \cdot 10^{38} $ & $ 0.64 $ &  $ 0.94 $ & $  83/67   $   \\     
NGC3377    & $ 0.4           $ &  $ 2.15 \pm 0.47 $ & $<1.8 \cdot 10^{37} $ & $ 3.0 \cdot 10^{38} $  & $ 3.3 \cdot 10^{38} $ & $ <0.06 $ &       -- & $   45/34  $   \\     
NGC3585    & $ 0.4           $ &  $ 2.75 \pm 0.61 $ & $<8.0 \cdot 10^{38} $ & $ 2.5 \cdot 10^{39} $  & $ 3.0 \cdot 10^{39} $ & $ <0.32 $ &       -- & $    46/39 $   \\     
NGC4278    & $ 0.47 \pm 0.06 $ &  $ 2.34 \pm 0.31 $ & $ 5.1 \cdot 10^{38} $ & $ 1.1 \cdot 10^{39} $  & $ 1.1 \cdot 10^{38} $ & $ 0.46 $ &  $ 0.52 $ & $ 151/132  $   \\     
NGC4365    & $ 0.49 \pm 0.04 $ &  $ 1.67 \pm 0.17 $ & $ 2.4 \cdot 10^{39} $ & $ 2.8 \cdot 10^{39} $  & $ 9.0 \cdot 10^{38} $ & $ 0.86 $ &  $ 1.26 $ & $  128/98  $   \\     
NGC4636    & $ 0.59 \pm 0.01 $ &  $ 2.64 \pm 0.05 $ & $ 1.1 \cdot 10^{41} $ & $ 1.3 \cdot 10^{40} $  & $ 4.1 \cdot 10^{38} $ & $ 8.46  $ &  $ 8.74$ & $ 1122/199 $   \\     
NGC4697    & $ 0.33 \pm 0.03 $ &  $ 2.54 \pm 0.54 $ & $ 7.3 \cdot 10^{38} $ & $ 4.7 \cdot 10^{38} $  & $ 2.1 \cdot 10^{38} $ & $ 1.55 $ &  $ 2.81 $ & $  78/66   $   \\     
NGC5128    & $ 0.62 \pm 0.01 $ &  $ -0.76 \pm 0.01$ & $ 2.1 \cdot 10^{39} $ & $ 3.0 \cdot 10^{40} $  & $ 5.1 \cdot 10^{37} $ & $ 0.07 $ &  $ 0.07 $ & $ 3547/435 $   \\     
Sagittarius& $ 0.4           $ &  $ 2.44 \pm 0.51 $ & $<1.2 \cdot 10^{33}$  & $ 2.5 \cdot 10^{33} $  &                    -- & $ <0.48 $ &  $ <0.48 $ & $  18/20   $   \\     
\hline \\
\end{tabular}
\end{minipage}
(1) Temperature of the thermal emission (2) Photon index of the power-law (3) and (4) Luminosities of the thermal and power-law components in the $0.5-8$ keV energy range (5) The combined luminosity of unresolved low-mass X-ray binaries brighter than $10^{35} \ \mathrm{erg \ s^{-1}}$  (but fainter than the sensitivity limit for the given  galaxy)  estimated from the K-band luminosity of the studied region and the average LMXB X-ray luminosity function of \citet{gilfanov} (6) The observed luminosity ratio of the soft thermal component to the power-law (7) Same as (6) but corrected for contribution of unresolved LMXBs. For NGC3377 and NGC3585, the predicted LMXB luminosity exceeds the observed luminosity of the power-law component, presumably  because of  a scatter in X/K ratios for LMXBs, therefore no LMXB-corrected luminosity ratio is computed. (8) Goodness of fit.
\label{tab:fit}
\end{table*}

It is known that the distribution of unresolved compact objects follows the stellar mass \citep{revnivtsev2,bogdan,revnivtsevm105}. Therefore, we looked for deviations in  the X-ray profile from the distribution of the K-band light as an indication of an  additional emission component,  presumably the emission of warm ionized gas. There are five galaxies, M32, M105, NGC3377, NGC3585, and Sagittarius in which the X-ray brightness closely follows the stellar light distribution at all central distances. In several others -- M31, M60, NGC1291, and NGC4278 the X-ray emission follows the  near-infrared light only in the outer regions. In the inner parts  of these galaxies, an additional X-ray emitting component is present and often dominates. In all other cases, the X-ray surface brightness strikingly deviates from the near-infrared light distribution, indicating the presence of strong additional X-ray emitting components. The largest difference between the X-ray and K-band profiles are observed in M84, NGC4636, and NGC5128.  These galaxies are known to show recent activity in their  nuclei \citep[e.g.][]{finugenov,kraft,baldi}.

To confirm that there is emission from ionized gas, we investigated the spectra of unresolved emission  (Fig. \ref{fig:spectra}).  As the gas emission is more centrally concentrated in some of the galaxies, we distinguished between  inner and outer regions. The dividing radii are listed in Table \ref{tab:list2}. Similar to the radial profiles, we excluded the contribution of resolved compact sources.  To facilitate the comparison, all spectra, shown in Fig. \ref{fig:spectra}, were normalized to the same K-band luminosity of $ L_{K} = 10^{11} \ \mathrm{L_{K,\odot}} $ and projected to a distance of $ 10 $ Mpc.  In the case of Sagittarius and NGC3377, we only show the spectrum of the entire galaxy due to the relatively low number of counts. The emission from the hot ISM reveals itself as a soft  component, clearly visible in the spectra of many of the galaxies. To  quantitatively characterize  its contribution, we performed fits to the spectra of unresolved emission (Table \ref{tab:fit}), using the MEKAL model in XSPEC to represent the emission from ionized gas  and a power-law spectrum for the contribution  of unresolved compact sources. The spectra were derived from full regions whose parameters are presented in the Table \ref{tab:list2}, the metal abundances for the thermal component were fixed at solar values \citep{anders}, and the hydrogen column density was fixed at the Galactic value \citep{dickey}. Such a simple model does not always describe the observed spectra well from the statistical point of view, as illustrated by the high $ \chi^2 $ values given in the last column of Table \ref{tab:fit}. However, it does describe the spectra with relative accuracy better than $\lesssim 10 \%$,  which is entirely sufficient for the purpose of this calculation.
The contribution of the hot ISM is, in principle, characterized by the ratio of luminosities of thermal and power-law components. The latter, however, includes the contribution of unresolved LMXBs, which may be dominant for galaxies with point source detection sensitivity that is too high, $\ga 10^{36} \ \mathrm{erg \ s^{-1}}$, making the luminosity ratios also depend on the sensitivity of the available Chandra data. To compensate for this, we estimated the contribution of unresolved LMXBs using the average LMXB X-ray luminosity function of \citet{gilfanov} and subtracted their contribution from the luminosity of the  power-law component. These corrected values are shown in the column labeled $L_{MKL}/(L_{PL}-L_{LMXB})$ in Table \ref{tab:fit}.

The spectral analysis results  presented in Table \ref{tab:fit} lead to the conclusions consistent with the brightness profile analysis.
As expected from surface brightness profiles, M32, M105, NGC3377, NGC3585, and Sagittarius show the same spectral properties at all central radii, they all have a rather weak soft component. In M31 and NGC4278, the significant difference between the spectra is that the luminous soft component is present only in the inner region, not the outer ones, suggesting that the hot gas is centrally concentrated \citep{bogdan}. In all other galaxies, the soft component dominates at all central radii and can be well-fitted with an optically-thin thermal plasma emission model with temperature in the range of $ kT \approx 0.3-0.8 $ keV, in good agreement with previous studies \citep{sarazin,irwin1,sivakoff,randall}. 

Based on radial profiles and spectral analysis, we conclude that the following seven galaxies are relatively gas-poor and may be suitable for our analysis: the bulge of M31, M32, M105, NGC3377, NGC3585, NGC4278, and Sagittarius. In all other cases, the signatures of recent activity of the galactic nucleus  and/or  large amount of hot gas make the galaxies unsuitable for our study.

\section{Results}

\subsection{Resolved supersoft sources}
\label{sec:sss}

The high bolometric luminosity, $\sim 10^{37}-10^{38} \ \mathrm{erg \ s^{-1}}$, means that some (generally speaking unknown) fraction of the nuclear-burning white dwarfs is detected by \textit{Chandra} as supersoft sources, despite the low color temperature of their emission. These sources obviously should be included in computing the final X/K ratios. 
To separate them from LMXBs, we used the spectral properties of compact sources.  The temperature of the hydrogen burning layer is in the range of $ T_{\mathrm{eff}} \sim 30-100 $ eV, but we conservatively included all resolved sources with hardness ratios corresponding to the blackbody temperature lower than $ kT_{\mathrm{bb}} <200 $ eV. We used  this method for all galaxies, except for M31, where we relied on the catalog of supersoft sources from \citet{distefano}. 
Because of the increased source cell size (Sect. \ref{sec:sample}), some of the sources are merged into one. This may compromise identification of supersoft sources, because some of them could be confused with harder sources and missed in our analysis.  To exclude this possibility, we repeated our analysis with a nominal source cell size and did not find any difference in the list of supersoft sources.

\begin{table*}[!ht]
\caption{X-ray luminosities in the $0.3-0.7$ keV band of various X-ray emitting components in gas-poor galaxies.}
\renewcommand{\arraystretch}{1.4}
\begin{minipage}{18cm}
\centering
\begin{tabular}{c c c c c c c}
\hline 
Name &  $L_{\mathrm{X,unres}} $   & $ L_{\mathrm{X,sss}} $ & $ L_{\mathrm{X,gas}} $ & $ L_{\mathrm{X}}/L_{\mathrm{K}} $ & $ (L_{\mathrm{X}}/L_{\mathrm{K}})_{\mathrm{corr}} $  & Age \\
& ($\mathrm{erg \ s^{-1}} $) & ($\mathrm{erg \ s^{-1}}) $ & ($\mathrm{erg \ s^{-1}} $) & ($\mathrm{erg \ s^{-1} \ L_{K,\odot}^{-1}} $) & ($\mathrm{erg \ s^{-1} \ L_{K,\odot}^{-1}} $) & (Gyr) \\ 

   &  (1) &  (2) & (3) & (4) & (5) &(6) \\ 
\hline 
M31 bulge  &  $ 1.1  \cdot 10^{38} $ &  $ 3.5  \cdot 10^{37} $ & $ 7.9 \cdot 10^{37} $ & $ (1.8 \pm 0.2) \cdot 10^{27} $ & $ (2.2 \pm 0.2) \cdot 10^{27} $ &6-10$^a$ \\ 
M32        &  $ 6.4  \cdot 10^{35} $ &  $ 9.0  \cdot 10^{35} $ & --                    & $ (1.8 \pm 0.2) \cdot 10^{27} $ & $ (2.4 \pm 0.2) \cdot 10^{27} $ &4-10$^b$ \\ 
M105       &  $ 6.4  \cdot 10^{37} $ &  $ 1.9  \cdot 10^{37} $ & --                    & $ (2.0 \pm 0.2) \cdot 10^{27} $ & $ (2.0 \pm 0.2) \cdot 10^{27} $ &8-15$^c$ \\
NGC3377    &  $ 6.7  \cdot 10^{37} $ &  $ 1.1  \cdot 10^{37} $ & --                    & $ (3.9 \pm 0.6) \cdot 10^{27} $ & $ (2.4 \pm 0.6) \cdot 10^{27} $ & 4.1$^d$ \\
NGC3585    &  $ 6.2  \cdot 10^{38} $ &          --             & --                    & $ (4.1 \pm 0.3) \cdot 10^{27} $ & $ (2.6 \pm 0.3) \cdot 10^{27} $ & 3.1$^d$ \\
NGC4278    &  $ 5.5  \cdot 10^{38} $ &  $ 3.4  \cdot 10^{37} $ & $ 4.1 \cdot 10^{38} $ & $ (3.2 \pm 0.2) \cdot 10^{27} $ & $ (3.2 \pm 0.2) \cdot 10^{27} $ &10.7$^d$ \\ 
Sagittarius&  $ 3.5  \cdot 10^{32} $ &  $ 7.3  \cdot 10^{32} $ & --                    & $ (1.2 \pm 0.4) \cdot 10^{27} $ & $ (1.7 \pm 0.4) \cdot 10^{27} $ &6.5-9.5$^e$\\ 
Milky Way  &                      -- &                      -- & --                    & $ (3.4 \pm 1.0) \cdot 10^{27} $ & --                              &--         \\ 
\hline \\
\end{tabular}
\end{minipage}
(1) Luminosity of unresolved X-ray emission (2) Luminosity of resolved supersoft sources (3) Luminosity of ionized gas (4) Observed X-ray to K-band ratio (5) X-ray to K-band ratio transformed to the same point source detection sensitivity of $ 2 \cdot 10^{36} \ \mathrm{erg \ s^{-1}} $. The errors in columns (4) and (5) correspond to statistical uncertainties in X-ray count rates. (6) Age of the stellar population. References for ages: $^a$ \citet{olsen}, $^b$ \citet{coelho}, $^c$ \citet{gregg}, $^d $ \citet{terlevich}, $^e$ \citet{bellazzini} \\
\label{tab:xtok}
\end{table*}  

\begin{figure}
\resizebox{\hsize}{!}{\includegraphics{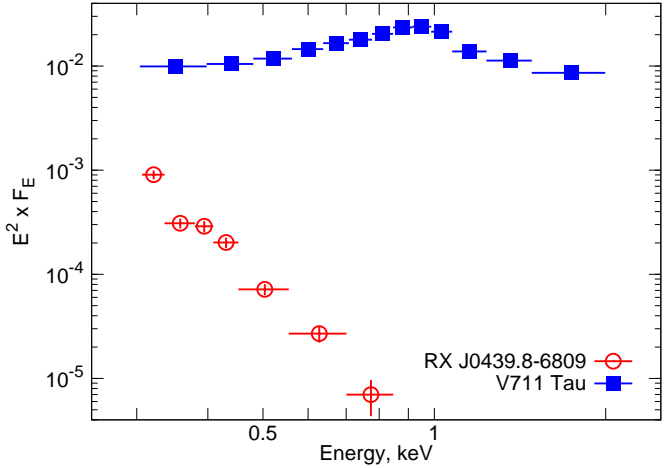}}
\caption{The spectrum of the active binary V711 Tau and of a steady nuclear-burning white dwarf of RX J0439.8-6809. The spectrum of the former was extracted using \textit{XMM-Newton} data, while the latter was observed by \textit{Chandra}. The background is subtracted from both spectra.}  
\label{fig:softsrc}
\end{figure}

\subsection{$L_X/L_K$ ratios}
\label{sec:xtokvalues}
The results of our analysis are presented in Table \ref{tab:xtok}. Listed in the table are the  X-ray luminosity of unresolved emission $L_{\mathrm{X,unres}} $, of resolved supersoft sources $ L_{\mathrm{X,sss}} $, and of ionized gas $L_{\mathrm{X,gas}}$. From these quantities and from the K-band luminosity of the studied region (Table \ref{tab:list2}), we computed the X-ray to K-band luminosity ratios  -- $L_{\mathrm{X}}/L_{\mathrm{K}}$. In the case of M31 and in NGC4278 we computed $L_X/L_K$ ratio in the outer regions, where no significant gas emission is detected. The Milky Way value was obtained using the results of \citet{sazonov}, who computed the luminosity of active binaries (ABs) and  cataclysmic variables (CVs) in the solar neighborhood.  The interstellar absorption is negligible in this case. We converted their X-ray-to-mass ratios from the $ 0.1-2.4 $ keV band to the $ 0.3-0.7 $ keV energy range, using typical spectra of ABs and CVs (see Fig. \ref{fig:softsrc}), and assuming a mass-to-light ratio of $ M_{\star}/L_{K} =1.0 $ \citep{kent}

The obtained ratios are in the range of $ L_{\mathrm{X}}/L_{\mathrm{K}} = (1.2-4.1) \cdot 10^{27} \ \mathrm{erg \ s^{-1} \ L_{K,\odot}^{-1}} $ and show a large dispersion. 
This dispersion is caused by the large difference in the point source detection sensitivity for the galaxies in our sample (Table 1).
To correct for this effect and to bring all galaxies to the same source detection sensitivity, we chose  the threshold luminosity of $ 2 \cdot 10^{36} \ \mathrm{erg \ s^{-1}} $. In those galaxies that had better source detection sensitivity  (M31 bulge, M32, M105, Sagittarius), we did not remove any source fainter than $ 2 \cdot 10^{36} \ \mathrm{erg \ s^{-1}} $ in computing the luminosity of ``unresolved'' emission. In the case of NGC3377 and NGC3585 having much worse detection sensitivity we subtracted the contribution of unresolved LMXBs in the luminosity range of $ 2 \cdot 10^{36} -  2 \cdot 10^{37} \ \mathrm{erg \ s^{-1}} $ from the measured luminosity of unresolved emission. To estimate the former we used  two methods. In the first, we measured the combined X-ray emission from hard resolved sources in this luminosity range in three galaxies, M31, M105, and NGC4278, which allowed us to compute the  $L_{\mathrm{X}}/L_{\mathrm{K}}$ ratio due to such sources for each of these three galaxies. We obtained fairly uniform values with the average number of $ \approx (1.5 \pm 0.2) \cdot 10^{27} \ \mathrm{erg \ s^{-1} \ L_{K,\odot}^{-1}} $, where the cited error is the rms of the calculated values. In the second method, the contribution of unresolved hard sources was estimated from the luminosity function of LMXBs \citep{gilfanov}. In the luminosity range of $ 2 \cdot 10^{36} -  2 \cdot 10^{37} \ \mathrm{erg \ s^{-1}} $, the X-ray to K-band luminosity ratio is $ \approx 1.4 \cdot 10^{28} \ \mathrm{erg \ s^{-1} \ L_{K,\odot}^{-1}} $ in the $2-10$ keV band. To convert this value to the $ 0.3-0.7 $ keV energy range we used the average spectrum of LMXBS, described by a power-law model with a slope of $ \Gamma = 1.56 $ \citep{irwin2} and assumed a column density of $ N_{H} = 4 \cdot 10^{20} \ \mathrm{cm^{-2}} $. The result is $L_{\mathrm{X}}/L_{\mathrm{K}} \approx (1.7\pm0.1) \cdot 10^{27} \ \mathrm{erg \ s^{-1} \ L_{K,\odot}^{-1}} $, which is in reasonable agreement with the value obtained from the first method.

Both methods are based on the assumption that the X/K ratio for LMXBs is the same for all galaxies in the sample. This assumption may be contradicted by the fact that in NGC 3377 and NGC 3585 the predicted luminosity of unresolved LMXBs exceeds the observed luminosity of unresolved emission (Table \ref{tab:fit}). Incidentally or not, these are the two youngest galaxies in our sample. The possible age dependence of the LMXB X/K ratio cannot be excluded but still needs to be established. On the other hand, the correction due to unresolved LMXBs is less than $\la 40\%$ of the observed value of $L_X/L_K$ (Table \ref{tab:xtok}).  This accuracy is sufficient for the present study, whose purpose is to constrain the luminosity of nuclear-burning white dwarfs. Therefore we defer further investigation of the possible effect of  inconstant LMXB X/K ratio  for a follow-up study.

The X-ray to K-band luminosity ratios  transformed to the same point source detection sensitivity  are listed in Table \ref{tab:xtok}. These numbers are fairly uniform $ L_{\mathrm{X}}/L_{\mathrm{K}} = (2.4 \pm 0.4) \cdot 10^{27} \ \mathrm{erg \ s^{-1} \ L_{K,\odot}^{-1}} $, where, as before, the cited error refers to the rms of the measured values. 

\subsection{Contribution of unresolved LMXBs}

The excellent source detection sensitivity, achieved in the bulge of M31, and the large number of compact X-ray sources in this galaxy allows us to estimate the contribution of unresolved LMXBs having the luminosities below the adopted threshold of $2\cdot 10^{36} \ \mathrm{erg \ s^{-1}}$   to the $L_X/L_K$ ratio. We consider the inner $ 6 \arcmin $ of the bulge where the source detection is complete down to $ 2 \cdot 10^{35} \ \mathrm{erg \ s^{-1}} $ \citep{voss}. In this region we collected all compact sources with the luminosity in the range $2 \cdot 10^{35} - 2\cdot 10^{36} \ \mathrm{erg \ s^{-1}}$, excluding those classified as supersoft sources. 
The combined  X-ray luminosity of these sources is divided by the near-infrared luminosity of the same region, to produce $L_X/L_K= (3.6 \pm 0.3) \cdot 10^{26} \ \mathrm{erg \ s^{-1} \ L_{K,\odot}^{-1}} $.  
This number represents the $L_X/L_K$ ratio in the soft band of low-mass X-ray binaries with luminosities in the $2 \cdot 10^{35} - 2\cdot 10^{36} \ \mathrm{erg \ s^{-1}}$  range.
We conclude that LMXBs contribute $ \sim 15 $ per cent to the  $L_X/L_K$ ratio derived above.

\subsection{The effect of the interstellar absorption}
\label{sec:absorption}

The possible effect of the interstellar absorption on the observed X-ray luminosities, depends on the energy spectra of the main X-ray emitting components -- active binaries and supersoft sources. ABs have significantly harder spectra than supersoft sources, which is illustrated in Fig. \ref{fig:softsrc}. The class of ABs is represented by V711 Tau, it was observed by \textit{XMM-Newton} for $ 3.2 $ ks in Obs-ID 0116340601, while RX J0439.8-6809 is an example of steady hydrogen-burning sources, based on a \textit{Chandra} exposure with $ 8.1 $ ks in  Obs-ID 83. As a consequence of the harder spectra, ABs are less affected by the interstellar absorption. 

\begin{figure}
\resizebox{\hsize}{!}{\includegraphics{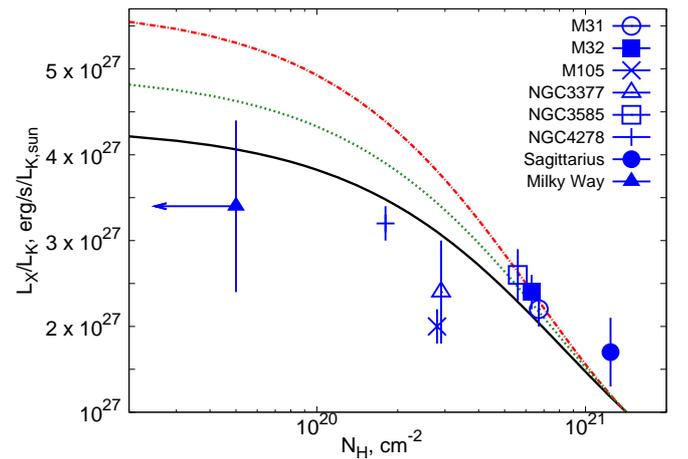}}
\caption{The sensitivity-corrected $L_X/L_K$ ratio versus Galactic hydrogen column density.  The solid line (black) shows the dependence of the absorbed flux on the $N_H$ assuming a power-law emission spectrum with slope $\Gamma=2$. The dotted line (green in the color version of this plot) and the dashed-dotted line (red) correspond to the combination of the same power-law with a blackbody spectrum with $kT_{bb}=75$ eV and $kT_{bb}=50$, respectively. For the last two curves equal luminosities of the two components in the $0.3-0.7$ keV band was assumed. Milky Way  represents the absorption-free $L_X/L_K$ ratio, and its $N_H$ value is shown as an upper limit and is arbitrarily placed along the \textit{x}-axis.}  
\label{fig:xtokplot}
\end{figure}

In Fig. \ref{fig:xtokplot} we plot the corrected X/K ratio ($\left(L_X/L_K\right)_{corr}$ in Table 3)  against the Galactic column density.  A weak anti-correlation between these two quantities appears to exist. This dependence or, rather, absence of a stronger one, can be used, in principle, to further constrain the contribution of sources with soft spectra to the X/K ratio. Indeed, the data is roughly consistent with the $N_H$ dependence for emission with a power-law spectrum with the slope  $\Gamma=2$. The value obtained for the solar neighborhood also fits  this dependence well. 
On the other hand, a much steeper dependence would be expected if a significant fraction (e.g. a half of the unabsorbed flux) of the $0.3 - 0.7$ keV emission had a blackbody spectrum with temperature of $50$ eV. 
This suggests that the contribution of sources with soft emission spectra, $kT_{bb}\sim 50-75$ eV is not dominant. The discrepancy decreases quickly, with the temperature of the soft emission and becomes negligible  for the blackbody temperature of  $kT_{bb}\sim 100$ eV, which has approximately the same dependence as a $\Gamma=2$  power-law. Therefore the contribution of the sources with harder spectra could not be constrained using this method. 

Because of the remaining dispersion in the data points caused by unknown systematic effects, the data cannot be adequately fitted in a strict statistical sense by a combination of power-law and soft spectral components with reasonable parameters. For this reason quantitative constraints on the contribution of sources with soft spectra would not be feasible. However, the qualitative conclusion from the analysis of the $N_H$ dependence of the $L_X/L_K$ ratio generally agrees with the result of \citet{sazonov}, who conclude that the emission from accreting white dwarfs contributes $ \lesssim 1/3 $ to the $L_X/L_K$ ratio of the solar neighborhood in the $ 0.1-2.4 $ keV energy band.

\subsection{Dependence of $L_X/L_K$ ratios on the parameters of galaxies }
\label{sec:age}

\begin{figure}
\resizebox{\hsize}{!}{\includegraphics{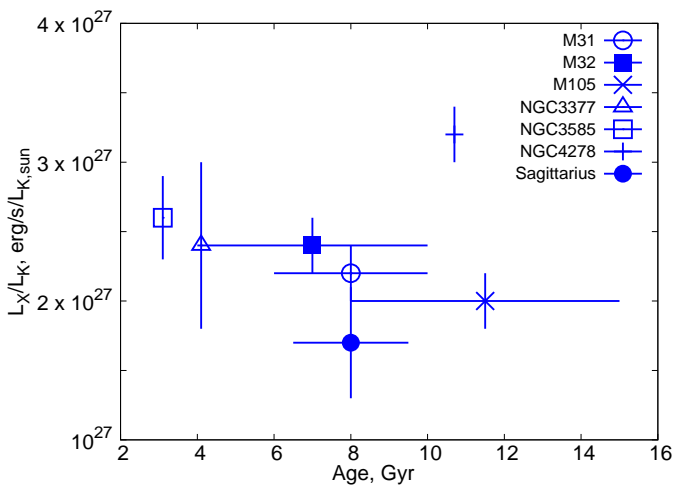}}
\caption{The sensitivity-corrected X-ray to K-band luminosity ratio versus the age of the stellar population.}  
\label{fig:agedistribution}
\end{figure}

We did not find any obvious correlations of the sensitivity-corrected $L_X/L_K$ ratios with the mass, metallicity \citep{terlevich}, and age (Fig. \ref{fig:agedistribution}) of the host galaxy. In particular,  we tentatively conclude that there is  no significant difference between younger ($\sim 3-4$ Gyrs) and older ($\sim 6-12$ Gyrs) early-type galaxies in our sample (but see the comment at the end of Sect. \ref{sec:xtokvalues} regarding the possible effect of the inconstant LMXB X/K ratio on the result of the sensitivity correction applied to the two youngest galaxies in our sample).

\section{Conclusions}

Using \textit{Chandra} archival data, we measured  the X-ray to K-band luminosity ratio in the $ 0.3-0.7 $ keV energy band in a sample of nearby gas-poor, early-type galaxies. 
In computing the X/K ratios, we retained only those components of X-ray emission that could be associated with the emission of steady nuclear-burning white dwarfs, namely unresolved emission and emission of resolved supersoft sources. To this end, we excluded  gas-rich galaxies from our sample and removed the contribution of resolved low-mass X-ray binaries.
Our final sample contains seven external galaxies covering a broad range of stellar masses and galaxy ages. It was complemented by the solar neighborhood data.

We measured a fairly uniform  set of X/K ratios with an average value of  $L_X/L_K=(2.4\pm0.4)\cdot 10^{27} \ \mathrm{erg \ s^{-1} \ L_{K,\odot}^{-1}}$. The error associated with this number corresponds to the rms of the values obtained for individual galaxies. We estimated that unresolved low-mass X-ray binaries contribute $\sim 15$ per cent of this value.
We did not find any significant dependence of X/K ratios on the parameters of the galaxies, such as their mass, age, or metallicity. There appears to be a weak anti-correlation of X/K ratios with the galactic absorption column $N_H$. The relative flatness of this dependence suggests that contribution of the sources with soft spectra $kT_{bb}\lesssim 50-75$ eV to this ratio does not dominate. The remaining dispersion in the data points precludes more rigorous and quantitative conclusions.  

\bigskip

\begin{small}
\noindent
\textit{Acknowledgements.}
We thank the anonymous referee for his/her useful and constructive comments. This research made use of \textit{Chandra} archival data provided by the \textit{Chandra} X-ray Center in the application package CIAO. \textit{XMM-Newton} is an ESA science mission with instruments and contributions directly funded by ESA Member States and the USA (NASA). This publication makes use of data products from Two Micron All Sky Survey, which is a joint project of the University of Massachusetts and the Infrared Processing and Analysis Center/California Institute of Technology, funded by the NASA and the National Science Foundation. The \textit{Spitzer Space Telescope} is operated by the Jet Propulsion Laboratory, California Institute of Technology, under contract with the NASA. 
\end{small}

\end{document}